\documentclass[prd,tightenlines,nofootinbib,superscriptaddress,twocolumn]{revtex4-1}
\usepackage[colorlinks=true,linkcolor=blue, urlcolor = blue ,citecolor=purple,linktocpage=true]{hyperref}

\usepackage{mathtools}
\usepackage{amssymb}
\usepackage{dsfont}
\usepackage{lipsum}
\usepackage{comment}
\usepackage{mathtools}
\usepackage{placeins}
\usepackage{xcolor}

\usepackage[makeroom]{cancel}
\usepackage{adjustbox}

\def\be{\begin{equation}}
\def\ee{\end{equation}}
\def\bea{\begin{eqnarray}}
\def\eea{\end{eqnarray}}
\def\beq{\begin{eqnarray}}
\def\eeq{\end{eqnarray}}

\def\nn{\nonumber}

\newcommand{\ii}{{\rm i}}
\definecolor{dblue}{rgb}{0, 0.25, 1}
\def\nn{\nonumber}
\def\dd{\mathrm{d}}
\usepackage{soul}

\newcommand{\Aa}{\mathcal{A}}

\newcommand{\bk}{{\mathbf k}}

\usepackage[stable]{footmisc}

\newcommand{\br}{{\bf{r}}}

\newcommand\ees{\end{eqnarray}}
\newcommand\bees{\begin{eqnarray}}

\newcommand{\cc}{\ \text{+ c.c.}}
\newcommand{\cm}{\ \text{+ c.m.}}

\definecolor{Mgreen}{rgb}{0.1, 0.69,0.16}

\begin{document}

\title{ Wave optics lensing of gravitational waves:  theory and phenomenology of triple systems in the LISA band}
\author{{\bf Martin Pijnenburg}}\email{martin.pijnenburg@unige.ch}
\affiliation{Département de Physique Théorique and Center for Astroparticle Physics, Université de Genève, Quai E. Ansermet 24, CH-1211 Genève 4, Switzerland}
\author{{\bf Giulia Cusin}}\email{giulia.cusin@iap.fr}
\affiliation{Institut d'Astrophysique de Paris, UMR-7095 du CNRS, Paris, France}
\affiliation{Département de Physique Théorique and Center for Astroparticle Physics, Université de Genève, Quai E. Ansermet 24, CH-1211 Genève 4, Switzerland}
\author{{\bf Cyril Pitrou}}\email{pitrou@iap.fr}
\affiliation{Institut d'Astrophysique de Paris, UMR-7095 du CNRS, Paris, France}
\author{{\bf Jean-Philippe Uzan}}\email{uzan@iap.fr}
\affiliation{Institut d'Astrophysique de Paris, UMR-7095 du CNRS, Paris, France}
\affiliation{Center for Gravitational Physics and Quantum Information,
Yukawa Institute for Theoretical Physics, Kyoto University, 606-8502,
Kyoto, Japan.}

\date{\today}
\begin{abstract}
We study lensing of gravitational waves by a black hole in the deep wave optics
regime, i.e. when the wavelength is much larger than the black hole Schwarzschild radius. We apply it to triple
systems, with a binary of stellar mass objects in the inspiraling phase orbiting around a central massive black
hole. We describe the full polarisation structure of the wave and derive predictions for the polarisation modes of the scattered wave measured by the observer. We show that lensing in the wave optics regime is not helicity preserving, as opposed to lensing in the geometric optics regime. The amplitude of the total wave is modulated due to interference between the directly transmitted and lensed components.
The relative amplitude of the modulation is fixed by the lensing geometry and can reach unity in the most favourable settings. This indicates that wave optics lensing is potentially detectable by LISA for sufficiently high SNR systems. Our findings show that in the wave optics regime it is necessary to go beyond the usual lensing description where the amplification factor is assumed to be the same for both helicity modes. 
While motivated by GW190521 and the AGN formation scenario, our results apply more
broadly to stellar-mass binaries orbiting a third body described as a Schwarzschild  black hole, with a period comparable to the GW observation time.
\end{abstract}
\maketitle




\section{Introduction}

Lensing of gravitational waves (GW) beyond geometric optics has been the subject of several studies in the last few years. Geometric optics is an approximation that holds when $2 M G/c^2 \gg \lambda$, where $M$ is the lens mass and $\lambda$ the wavelength of the wave. While this condition is always satisfied when considering lensing of GW  in the frequency band of ground-based detectors, beyond geometric-optics effects are expected to appear in the frequency band of the Laser Interferometer Space Antenna (LISA). 

The existing literature on lensing beyond geometric optics  can be divided into two categories. Refs.\,\cite{Takahashi:2003ix, Tambalo:2022plm, Savastano:2023spl} 
relax the geometric optics assumption and consider the regime in which the GW wavelength is either comparable or larger than the lens size. However, they still neglect the spin-nature of the wave, effectively treating it as a scalar wave. 
On the other hand, there are studies of GW lensing beyond geometric optics that do not neglect the spin-2 nature of the wave, either introducing a perturbative approach \cite{Cusin:2019rmt, Dalang:2021qhu, Harte:2018wni, Dolan:2018ydp} or importing quantum scattering techniques, see Refs.~\cite{futterman_handler_matzner_1988} for a review and~\cite{Dolan:2007ut} for a more recent work on the subject. 

It is however not straightforward to compute the phenomenology of realistic situations from these formal results.

In this article, we present a framework to study lensing of 
GW in the deep wave optics regime, i.e.  when the wavelength is much larger than the black hole Schwarzschild radius. 
 We keep track of the full tensorial structure of the wave and we develop an approach to apply our results to realistic situations. 
We stress that GW lensing in wave optics regime is not polarisation  preserving, 
i.e. the ratio of the two helicity modes is not conserved during propagation, unlike in the geometric optics limit. We derive the transformation properties of the two polarisation modes under lensing, showing that helicity modes are mixed, in contrast with what is usually assumed in the literature when studying GW lensing with a common amplification factor both polarisation modes  (see e.g. Ref.~\cite{Takahashi:2003ix}). 
We then apply our finding to study lensing phenomenology of triple systems, in which a binary of compact objects in the inspiralling phase orbits around a central massive black hole. We show that interference between the scattered and transmitted components of the wave gives rise to a modulation of the total waveform, potentially detectable by LISA.

Our study is motivated by the features of black hole merger GW190521, measured by LIGO and Virgo during the last observing run, which suggest the binary might have undergone stronger than expected interactions with its environment. 
GW190521 had component masses  
$85^{+21}_{-14}\,M_\odot$ and $66^{+17}_{-18}\, M_\odot$~\cite{Abbott:2020tfl, Abbott:2020mjq}, 
with the larger lying within the pair-instability gap $\sim[50,130] M_\odot$ \cite{2002RvMP...74.1015W,2003ApJ...591..288H,2019ApJ...887...53F}.
A possibility is that the progenitor black holes (BHs) may have formed via a ``dynamical'' formation channel, i.e. by repeated coalescences in dense environments such as 
 globular or nuclear stellar clusters \cite{2017PhRvD..95l4046G,2019PhRvD.100d3027R,2020arXiv200905065F}
or  active galactic nuclei~(AGNs)~\cite{2007MNRAS.374..515L,2020MNRAS.494.1203M,2020ApJ...898...25T}.  
Moreover, the Zwicky Transient Facility (ZTF) detected
an optical flare (dubbed ZTF19abanrhr) about 34 days
after GW190521, in AGN J124942.3+344929 at redshift $z = 0.438$. The position and distance of this system are compatible with
the inferred position and distance of GW190521. 
This event was then interpreted as due to the BH remnant from GW190521
moving in the AGN disk,  Ref.~\cite{PhysRevLett.124.251102}, as a result of the recoil produced
by the anisotropic gravitational wave emission during the merger. 
Reference~\cite{PhysRevLett.124.251102} argued that the distance of 
the GW190521 binary from 
the nucleus should be about $700 GM/c^2$, with $M\sim10^8 -10^9 \ M_{\odot}$ the mass of the  BH at the center of the AGN.

If indeed GW190521 lived in an AGN disk, it could belong to a significant population of binary BHs located in dense environments that will be detected in the coming years by ground and space detectors~\cite{Barausse:2007dy,Barausse:2014tra,Caputo:2020irr}. 
 The presence of the central super massive black hole can significantly affect the waveform, as a result of the accelerated motion of the stellar-origin BH binary around it~\cite{Bonvin:2016qxr,Inayoshi:2017hgw,Tamanini:2019usx}, or because of lensing~\cite{DOrazio:2019fbq,2018MNRAS.474.2975D, Toubiana:2020drf, Sberna:2022qbn} and Shapiro time delay~\cite{1964PhRvL..13..789S, Toubiana:2020drf, Sberna:2022qbn}. 
 Several of these environmental effects could be observable with LISA, by targeting the early inspiral of stellar-origin BH binaries months/years before they merge in the band of ground detectors~\cite{Inayoshi:2017hgw,Tamanini:2019usx,Randall:2019sab,Hoang:2019kye,Deme:2020ewx,Caputo:2020irr,Yu:2020dlm,Toubiana:2020cqv,Toubiana:2020drf}.\footnote{These effects appear at negative post-Newtonian (PN) orders in the GW phase relative to the vacuum quadrupole emission~\cite{Barausse:2014tra,Toubiana:2020drf}, hence they are expected to be relevant at low frequencies.}

In Refs.~\cite{DOrazio:2019fbq,2018MNRAS.474.2975D, Toubiana:2020drf, Sberna:2022qbn}, lensing of such systems is described importing results developed to study lensing of electromagnetic radiation in the geometric optics regime, and treating the wave as a scalar  object. The condition defining the regime of validity of wave optics can be written as 
\be\label{waveoptics}
\left(\frac{M}{3\times 10^{7} M_{\odot}}\right)\left(\frac{f}{\text{mHz}}\right) \ll 1\,.
\ee
It follows that while for the particular system of GW190521, geometric optics is a good approximation of lensing phenomena, wave optics lensing has to be considered for a broader range of triple systems. 

In our phenomenological illustration, we consider a binary system with chirp mass $80M_\odot$ in a scenario where the lens is located at the low end of the AGN redshift distribution \cite{CAIXAcatalogueAGN}. The radius of the circular orbit of the binary around the central lens is chosen to be $662 GM/c^2$, following a numerical prescription by \cite{Bellovary:2015ifg} \footnote{Note that with a chirp mass of $80M_\odot$, the system has  massive components reaching the intermediate mass gap. However, AGN migrations traps have been found to be a favorable environment for the production of black holes in this mass range \cite{Bellovary:2015ifg}, making the assumption reasonable.}. 

We take $f=3\times10^{-3}$Hz to be within the LISA band, and consider an AGN mass of $M = 1.2\times10^6 M_\odot$. Such an AGN mass is not unrealistic as is shown in \cite{NGC4051}. Such a system lies well inside the undulatory regime (\ref{waveoptics}) and, if detected by LISA, the wave optics modulation leaves an observable signature on the waveform.

\begin{figure}[ht!!!]
\centering
\includegraphics[width = \linewidth]{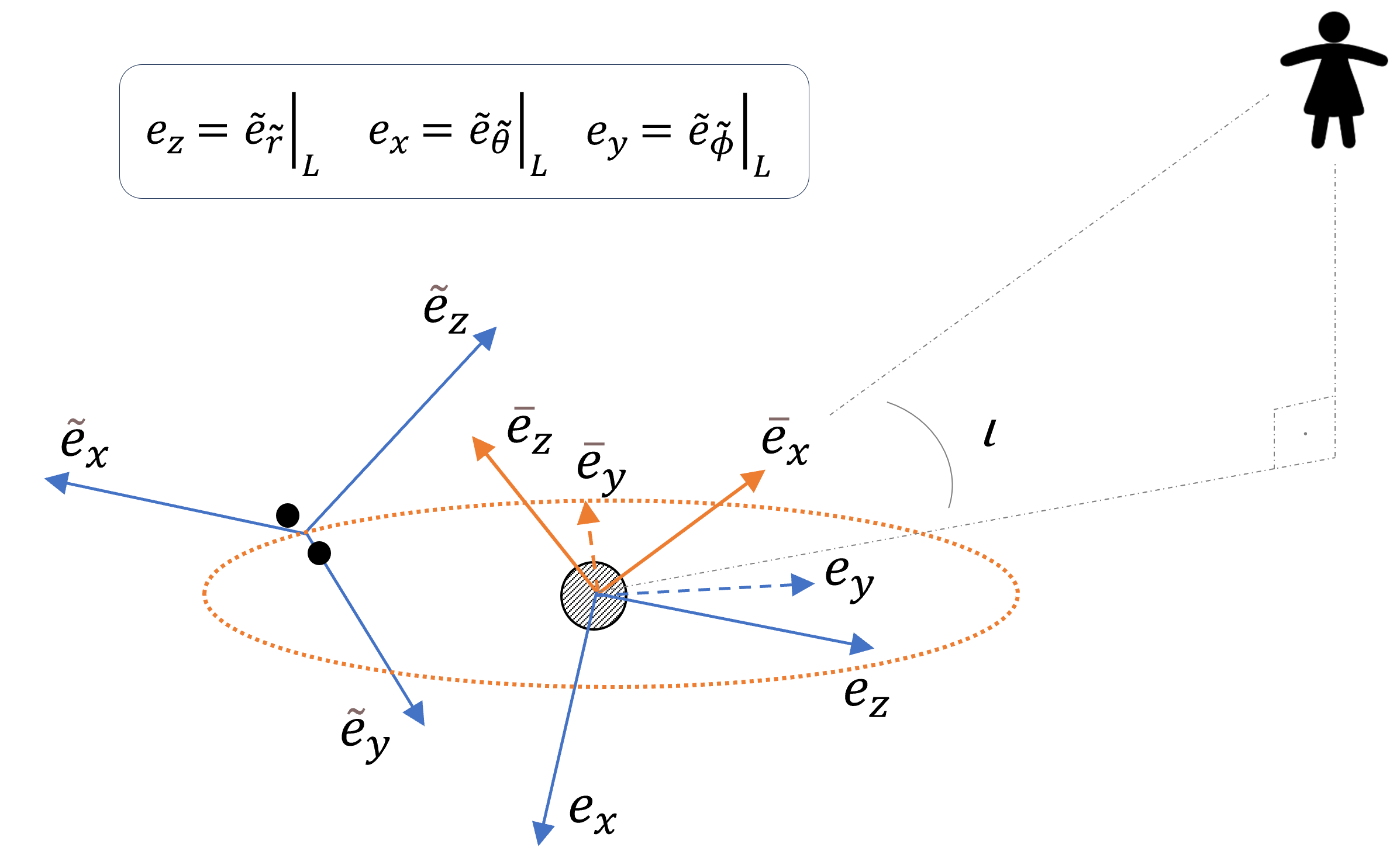}
\caption{A binary in the inspiraling regime orbits around a massive black hole, with $\tilde{e}_z$ along the binary's angular momentum. Waves emitted by the binary scatter off the third body's potential and reach the observer. Freezing the orbital motion, the static geometry (in blue) is described using the $\tilde{e}_{x, y, z}$ basis, naturally associated with the binary, and the derived $e_{x, y, z}$ naturally associated with the incoming wave on the third body. Adding the effect of orbital motion (orange) requires the introduction of a third set $\bar{e}_{x, y, z}$ from which $\tilde{e}_{x, y, z}$ may be related via Euler rotations of angles $\alpha_{1, 2, 3}$.}
\label{MovingScheme}
\end{figure}

In Fig.~\ref{MovingScheme} we depict the type of systems that we consider, and whose geometrical configurations are introduced step by step in the subsequent sections.

We stress that, while motivated by GW190521 and the AGN formation scenario, our results apply more broadly to stellar-mass binaries orbiting a third body with a period comparable to the GW observation time, \emph{in the absence of gas} (if matter is present around the black hole, additional effects are in principle present and have to be included in the waveform description, see e.g. \cite{Barausse:2014pra} for a review). 

The article is structured as follows: in Section~\ref{Summary} we summarize the method followed to construct the wave at the observer, in a static situation in which a binary system emits a spherical wave lensed by a black hole. In Section~\ref{sec_in} we explicit the wave emitted by the source (and that we assume to be described by the quadrupole formula) in a reference frame centered on the lens, and in terms of Regge-Wheeler variables. In Section~\ref{sec_scatt} we construct the scattered wave and show how the two helicity modes of the scattered wave (at the observer) are related to the emitted ones. In Section~\ref{sec_cross} we compute the cross section of this process, showing that we recover known results of the literature when taking proper limits. Finally, in Section~\ref{sec_pheno} we consider the dynamical situation in which the binary system rotates around the central back hole and we discuss phenomenological implications of lensing. In Section~\ref{sec_discussion} we discuss our findings and present our conclusions. Technical derivations are presented in a series of appendices. Table~\ref{table_summary} lists the geometrical quantities introduced in this work, and provides definitions. In this article, we work in natural units in which $G=c=1$.

\begin{table*}[htb!!!]
  \centering
  \begin{adjustbox}{width=\textwidth}
		\begin{tabular}[c]{ c | l }
				\hline
				$\tilde{e}_{x, y, z}$& \quad Cartesian basis vectors centered on the source, with $\tilde{e}_{z}$ along the source's internal angular momentum \\
				$(\tilde{\theta}, \tilde{\phi}, \tilde{r})$& \quad spherical coordinates associated to the Cartesian system $\tilde{e}_{x, y, z}$ \\ 
				$(\tilde{\theta}_X, \tilde{\phi}_X)$& \quad angular spherical coordinates of object $X$ w.r.t. $\tilde{e}_{x, y, z}$, for $X=L$ (lens) or O (observer) \\
				$\tilde{e}_{\tilde{\theta}, \tilde{\phi}, \tilde{r}}$& \quad spherical unit basis vectors of the system of spherical coordinates associated to $\tilde{e}_{x, y, z}$ \\
				$e_{x, y, z}$& \quad Cartesian basis vectors centered on the lens, defined as $e_{x, y, z} \equiv \tilde{e}_{\tilde{\theta}, \tilde{\phi}, \tilde{r}}|_{(\tilde{\theta}_L, \tilde{\phi}_L)}$ \\
				$(\theta, \phi, r)$& \quad spherical coordinates associated to the Cartesian system $e_{x, y, z}$ \\ 
				$(\theta_\text{O}, \phi_\text{O})$& \quad angular spherical coordinates of the observer w.r.t. the lens's $e_{x, y, z}$ \\
				$e_{\theta, \phi, r}$& \quad spherical unit basis vectors of the system of spherical coordinates associated to $e_{x, y, z}$ \\
				$\eta$& \quad angle between $e_{\theta}|_\text{O}$ and $\tilde{e}_{\tilde{\theta}}|_\text{O}$\\
				\hline
				\hline 
				$\bar{e}_{x, y, z}$& \quad fixed Cartesian basis vectors centered on the lens, with $\bar{e}_{x}$ along the observer direction \\
				$\iota$ & \quad inclination angle of the outer orbital plane w.r.t. the source-lens axis \\ 
				$\alpha_{1, 2, 3}$ & \quad Euler angles that allow one to rotate the triad 	$\bar{e}_{x, y, z}$ into $\tilde{e}_{x, y, z}$\\
				\hline
		\end{tabular}
  \end{adjustbox}
  \caption{\label{table_summary}We summarise our notation, for the static (upper part) and circular orbit (full table) settings.} 
   \label{TableBig}
\end{table*}
\section{Summary of our approach}\label{Summary}

When studying the lensing of GWs off black holes in the wave optics regime, it is convenient to express the strain in terms of Regge-Wheeler variables. When using a reference frame $e_{x, y, z}$ centered on the lens (with the polar axis aligned with the lens-source direction), these variables satisfy a Schr\"odinger-like scattering equation, describing Rutherford scattering of a charged particle off a Coulomb-like potential. Then one can import techniques developed in the context of quantum scattering to compute phase shifts that characterise the deviation of the scattered wave with respect to an incoming plane wave.
While this analogy is apparent at a formal level, Rutherford scattering and classical scattering of GW off black holes share some common features: in both cases the potential associated to the target object is Coulomb-like. Moreover, a charged particle is electromagnetically charged in the same way as the gravitational wave travelling in the black hole space time is gravitationally charged and feel the gravitational potential of the lens \cite{Rutherford2022}. 
\medbreak
On the other hand the waveforms emitted by astrophysical objects are conveniently expressed in terms of plus and cross polarisations of the strain in the transverse-traceless (TT) gauge, since this allows one to easily compute the effect on detectors. Moreover, the waveforms emitted by compact binaries, are also more conveniently expressed using a  system of axes $\tilde{e}_{x, y, z}$ centered on the source, with polar axis normal to the orbital plane of the binary.
Therefore, the frame used to study GW scattering and the one used in waveform modelling are different, and we must relate them. 
\medbreak
Assuming that the source emits a monochromatic signal well described by the quadrupole formula (in TT gauge), we need to:
\begin{enumerate}
    \item assume that locally at the lens the wave front can be described as a plane wave and write the wave in a frame $e_{x, y, z}$ centered on the lens and with polar axis aligned with the source-lens direction. This leads to Eq.~\eqref{H};
    \item rewrite this plane wave using Regge-Wheeler variables associated to a multipolar decomposition, see Eqs.~\eqref{PsiEvenNoScattForm} and \eqref{PsiOddNoScattForm};
    \item compute the scattered wave using phase shifts results derived in the literature, i.e. Eqs.~\eqref{PhaseShiftMinus} and \eqref{dd1};
    \item write the final scattered wave in TT gauge in terms of plus and cross polarisation, leading to Eqs.~\eqref{hplScattered} and \eqref{hxScattered};
    \item reconstruct the total wave at the observer, in real space (scattered plus transmitted) in order to obtain observable lensing signatures. 
\end{enumerate}
\begin{figure}[ht!!!]
\centering
\includegraphics[trim=1.5cm 1.25cm 8.5cm 0, clip, width=1\linewidth]{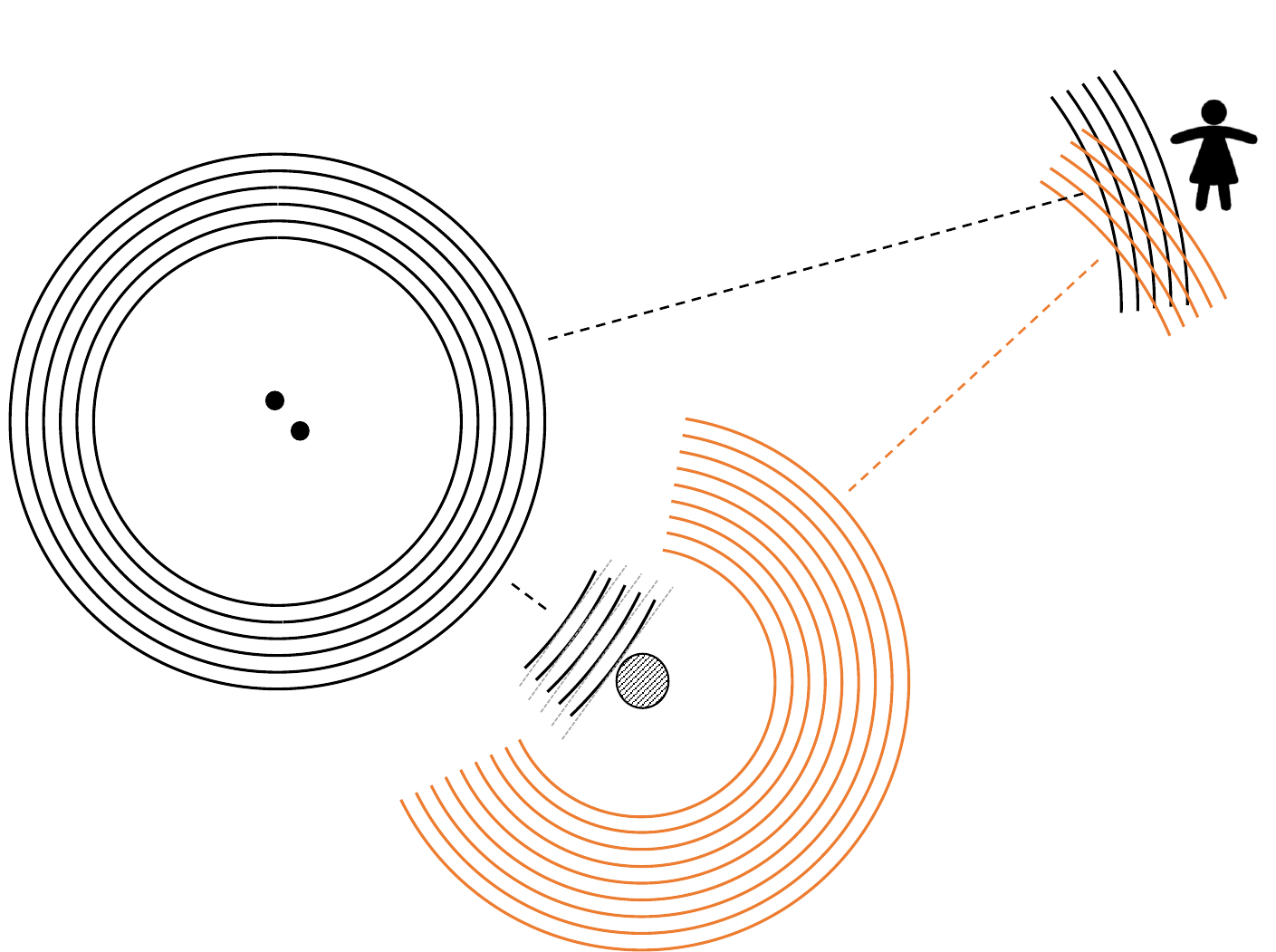}
\caption{Two dimensional schematic visualisation of the wave lensing as a scattering process: a spherical wave that is locally plane scatters off a third body, producing a scattered spherical wave centered on the latter. This scattered wave interferes with the incident one.}
\label{WaveDecompositionScheme}
\end{figure}
This description of wave optics lensing as a scattering phenomenon is summarised in Fig.~\ref{WaveDecompositionScheme}. 
In the remaining part of this article, we will go through these various steps, and the notation introduced is summarized in Table \ref{TableBig}. We start with a review of results on scattering of GWs off black holes in Regge-Wheeler variables. We first consider the static situation, in which the geometry lens-source-observer is fixed, and we then move to the time-dependent situation, in which the source is in orbit around the central lens. 

\subsection{Regge-Wheeler variables}

Let us consider a plane GW with wave vector $\bk$ as a perturbation of a Schwarzschild black hole. We choose the system of coordinates such that the $z$ axis is along $\bk$, and $|\bk|=\omega$. The background metric is given by 
\be
\dd s^2 = g_{ab} \dd x^a \dd x^b + r^2 \Omega_{AB}\dd x^A \dd x^B\,,
\ee
where $a,b=t, r$, $A,B = \theta,\phi$, the background metric on the 2-sphere is $\Omega_{AB}=\text{diag}(1, \sin^2\theta)$, and 
\be
g_{ab} \dd x^a \dd x^b=-\mathcal{B}(r) \dd t^2+\mathcal{B}(r)^{-1} \dd r^2\,,
\ee
with $\mathcal{B}(r)\equiv 1-2M/r$. 

Linearising the Einstein equations around this Schwarzschild background leads to a set of perturbation equations. It is standard lore  to recast them into two “master equations”: the Regge–Wheeler and the Zerilli equations for respectively the Regge–Wheeler function and the Zerilli  “master functions”, each containing modes of a given parity (odd and even, respectively); see Ref.~\cite{10.1093/oso/9780198570899.001.0001} for a pedagogical derivation. In vacuum, the Regge-Wheeler and Zerilli functions ($\Psi_{\text{odd}}$ and $\Psi_{\text{even}}$, respectively) satisfy the Schr\"odinger-like equation 
\be
\frac{\dd^2\Psi_{\bullet}}{\dd r_*^{2}}+\left(\omega^2-V_{\bullet}\right) \Psi_{\bullet}=0\,,
\label{ReggeWheeler}
\ee
where $\bullet=\left(\text{even, odd}\right)$ and  $\dd r_*/\dd r=1/\mathcal{B}$ defines the tortoise coordinate once we integrate it to 
\be
r_*(r) = r-2M\ln\left(\frac{r}{2M}-1\right)\,.
\label{tortoise}
\ee
The choice of integration constant above is made such that it is compatible with the phase shifts of \cite{PoissonSasaki}.
For the odd (axial) mode the potential is 
\be
V_{\text{odd}}=\mathcal{B}(r)\left[\frac{\ell(\ell+1)}{r^2}-\frac{6 M}{r^3}\right]\,,
\ee
and for the even (polar) one
\be
V_{\text{even}}=\frac{2\mathcal{B}(r)}{r^3}\frac{9 M^3+3\beta_{\ell}^2 M r^2+\beta_{\ell}^2(1+\beta_{\ell})r^3+9 M^2\beta_{\ell} r}{(3 M+\beta_\ell r)^2}\,,
\ee
with $\beta_\ell=(\ell-1)(\ell+2)/2$. In principle, the full perturbed metric in the Regge-Wheeler gauge can be reconstructed from these two master functions, and it can then be expressed in TT gauge.

\subsection{Gravitational wave scattering: phase shifts}
\label{sec:PhaseShift}

In direct analogy with quantum scattering theory, the differential equation \eqref{ReggeWheeler} can be solved such that the solution reduces to the original plane wave in absence of scattering ($M=0$). The scattered part of the solution can be described using a phase shift approach. The initial incoming wave is expressed as a superposition of spherically ingoing and outgoing waves, and the scattering process induces an extra outgoing spherical (scattered) wave. The effect of scattering can be recast into the ratio of the in- and outgoing spherical wave amplitudes (the former contains only the incoming wave contribution while the latter contains both incoming and scattered wave contributions). For a scalar initial plane wave $e^{\ii \bk\br}$, the general expansion of the wave $\psi$ is
\begin{widetext}
	\begin{align}\label{planewaveexp}
		& \psi\simeq \frac{\ii}{ 2\omega r}\sum_{\ell=2}^{\infty} (-1)^{\ell}(2\ell+1) \left[e^{-\ii \omega r_*}-(-1)^{\ell} e^{\ii\omega r_*} S_{\ell}(\omega)\right] P_{\ell}(\cos\theta)\,,
	\end{align}
\end{widetext}
 with $|\bk|=\omega$, $\bk\parallel z$ and 
 \be
 S_{\ell}(\omega)\equiv \hbox{e}^{2 \ii \delta_{\ell}(\omega)}\,.
 \ee

This set of functions describes the departure of the wave function from the plane wave, i.e. for $\delta_\ell=0$ the wave function coincides with the plane wave (asymptotic limit of the partial waves expansion). 

The gravitational case which we consider is analogous, with some caveats. First, the wave is not scalar but tensorial, so the partial wave expansion is not as straightforward as in Eq.~\eqref{planewaveexp}, since the expansion of a tensor valued wave has more multipolar indices ($\ell, m, s$). In place of the gravitational wave itself, the scattering process is conveniently phrased in terms of the scalar valued functions Regge-Wheeler and Zerilli $\Psi_\bullet$. This comes at the price of needing to translate the initial conditions on the locally plane metric perturbation into conditions on $\Psi_\bullet$. Hence we must first relate $\Psi_\bullet$ and the GW, a procedure built up from Ref.~\cite{Martel:2005ir} and detailed in \S~\ref{Martel}, where we show that $\Psi_\bullet$ can still be expanded as a superposition of incoming and outgoing spherical waves, thus allowing us to use the phase shift formalism.

In the $(\omega M)\ll 1$ long wavelength limit, analytic forms for the phase shifts associated with Eq.~\eqref{ReggeWheeler} have been found by Poisson and Sasaki \cite{PoissonSasaki}, thus solving the differential equation. They read:
 \begin{align}
 \label{PhaseShiftMinus}
 &\exp(2 \ii \delta_{\ell}^{\text{odd}})=\\
 &=\hbox{e}^{-\ii \Phi-2i M \omega}\hbox{e}^{8 \ii M\omega/\ell(\ell+1)}\frac{\Gamma(\ell+1-2 \ii M\omega)}{\Gamma(\ell+1+2\ii M\omega)}+\mathcal{O}(M^2\omega^2)\,,\nn
 \end{align}
with $\Phi=-4 M \omega \ln(4 M\omega)$. The phase shifts of even parity are related to those of odd parity by 
 \begin{align}\label{dd1} 
&\exp(2 \ii \delta_{\ell}^{\text{even}})=\nn\\
 &=\frac{(\ell+2)(\ell+1)\ell(\ell-1)+12 \ii M \omega}{(\ell+2)(\ell+1)\ell(\ell-1)-12 \ii M \omega} \exp(2 \ii \delta_{\ell}^{\text{odd}})\,.
 \end{align}
Note that these phase shifts are independent of $m$ since the potential is spherically symmetric.
The solution of Ref.~\cite{PoissonSasaki} shows that in the phases of the spherical wave modes, the radial coordinate $r$ has to be substituted by the tortoise $r_*$ that appears in the radial derivative of the differential equation \eqref{ReggeWheeler}. The appearance of logarithmic radial corrections in the phases is a generic feature of $1/r$ potentials and also appears in quantum Coulomb scattering \cite{Rutherford2022}.
 With this, we have all the ingredients to reconstruct scattering solutions for $\Psi_\bullet$, in the long wavelength limit.

\section{Incoming wave}\label{sec_in}

We consider the simple static setup, as in Fig.~\ref{MovingScheme}. The source emits a spherical wave, well described by the quadrupole formula in a system of axes centered on the source. We then rewrite this wave in a reference system centered on the lens, and in terms of Regge-Wheeler variables (instead of $+$ and $\times$ polarisations). 
We will assume that locally (around the lens) the spherical wavefront can be approximated with a plane wavefront, i.e. we treat the
wave impacting on the lens as a plane wave.

\subsection{Metric perturbations and Regge-Wheeler variables}\label{Martel}

Ref.~\cite{Martel:2005ir} provides the relations between Regge-Wheeler variables and the metric perturbations expanded in terms of even parity and odd parity spherical harmonics for 
a generic perturbation $h_{\mu\nu}$ of the Schwarzschild metric as

\begin{align}
h_{ab}&=\sum_{\ell m} h_{ab}^{\ell m}Y^{\ell m}\,,\label{pAB}\\
h_{aB}&=\sum_{\ell m}\left(j_a^{\ell m}Y_B^{\ell m}+h_a^{\ell m}X_B^{\ell m}\right)\,,\\
h_{AB}&=\sum_{\ell m}\left(r^2 K^{\ell m}\Omega_{AB}
Y^{\ell m}+r^2 G^{\ell m}Y_{AB}^{\ell m}+h_2^{\ell m}X_{AB}^{\ell m}\right)\,,\label{pABf}
\end{align}
where $Y^{\ell m}$ are the standard spherical harmonics, while $Y^{\ell m}_A$ (resp. $X^{\ell m}_{A}$) are even (resp. odd) vector spherical harmonics and $Y^{\ell m}_{AB}$ (resp. $X^{\ell m}_{AB}$) are even (resp. odd) tensorial spherical harmonics. These are functions on the sphere with given parity, whose relation to spin-weighted spherical harmonics is outlined in Appendix~\ref{matching}. The expansion factors $h^{\ell m}_{ab},h^{\ell m}_a,j^{\ell m}_a, K^{\ell m},G^{\ell m},h^{\ell m}_2$ are functions which depend on $r$ and $t$. 

The asymptotic identification of the even and odd Regge-Wheeler and Zerilli gauge invariant potentials is then found from gauge invariant combinations of metric perturbations, once the functions
$h^{\ell m}_{ab},h^{\ell m}_a,j^{\ell m}_a, K^{\ell m},G^{\ell m},h^{\ell m}_2$
are known. 
For instance in the odd case, with appendix C of \cite{Martel:2005ir}, we find 
in terms of gauge invariant quantities (denoted with a hat) 
\begin{equation}\label{Psioddfromhr}
\Psi_{\rm odd}^{\ell m} =\frac{2 r}{\mu}\left(\frac{\partial}{\partial r}\hat{h}^{\ell m}_t-\frac{\partial }{\partial t}\hat{h}^{\ell m}_r-\frac{2}{r}\hat{h}^{\ell m}_t\right)\,, 
\end{equation}
where $\mu=(\ell-1) (\ell+2)$. These gauge invariant variables are defined as
\begin{align}
\hat{h}^{\ell m}_t&=h^{\ell m}_t-\frac{1}{2}\partial_t h^{\ell m}_2\,,\\
\hat{h}^{\ell m}_r&=h^{\ell m}_r-\frac{1}{2}\partial_rh^{\ell m}_2+\frac{1}{r} h^{\ell m}_2\,.
\end{align}
For the even parity we have (see appendix C of Ref.~\cite{Martel:2005ir})
\be\label{Psioddfromhr_even}
\Psi_{\rm even}^{\ell m}=\frac{2r}{\mu+2}\left[\hat{K}^{\ell m}+\frac{2 \mathcal{B}}{\Lambda}\left( \mathcal{B} \hat{h}^{\ell m}_{rr}-r \frac{\partial}{\partial r} \hat{K}^{\ell m}\right)\right]\,,
\ee
where $\Lambda=\mu+6 M/r$
and where the gauge invariant quantities are defined as
\begin{align}
    \hat{h}^{\ell m}_{rr}&=h^{\ell m}_{rr}-2 \partial_r j^{\ell m}_r-\frac{2M}{r^2 \mathcal{B}}j^{\ell m}_r+r^2 \partial_r^2G^{\ell m}\nonumber\\ &+\frac{2r-3M}{
    \mathcal{B}}\partial_rG^{\ell m}\,,\\
    \hat{K}^{\ell m}&=K^{\ell m}-\frac{2\mathcal{B}}{r} j^{\ell m}_r+ r \mathcal{B}\partial_rG^{\ell m}+\frac{\mu+2}{2}G^{\ell m}\,.
\end{align}

\subsection{Quadrupole formula in the lens frame}

The source emits a spherical wave, described by the quadrupole formula. 
For any fixed lens angular position $(\tilde{\theta}_L, \tilde{\phi}_L)$ in the source sky, we take as basis vectors $(e_x, e_y, e_z)$ = $(\tilde{e}_{\tilde{\theta}}, \tilde{e}_{\tilde{\phi}}, \tilde{e}_{\tilde{r}})|_{(\tilde{\theta}_L, \tilde{\phi}_L)}$, where $(\tilde{e}_{\tilde{\theta}}, \tilde{e}_{\tilde{\phi}}, \tilde{e}_{\tilde{r}})$ are the spherical basis vectors of the source sky, at the lens position. 
The metric, in this $(\tilde{\theta}_L, \tilde{\phi}_L)$ 
direction can be written as 
\be
 h_{ij}=\left(
\begin{array}{ccc}
h_+&h_\times&0\\
h_\times&-h_+&0\\
0&0&0\\
\end{array}
\right)_{ij}\,,
\label{TTwaveMatrix}
\ee
in the $(e_x, e_y, e_z)$ Cartesian coordinate system, for a generic lens direction $(\tilde{\theta}_L, \tilde{\phi}_L)$. For a binary system of compact objects of masses $m_{1,2}$, the chirp mass is
\be
M_c = \frac{(m_1m_2)^{3/5}}{(m_1+m_2)^{1/5}}\,.
\ee
When the binary is in the inspiraling phase, which we describe to lowest order by a circular motion, the emitted wave in the lens (L) direction is \cite{Maggiore} (note our different convention for spherical coordinates w.r.t. that reference) 

\begin{align}
\label{PlusRealExpression}
h_+(t, \tilde{r}, \tilde{\theta}_L, \tilde{\phi}_L) &=  \frac{A_\text{in}}{\tilde{r}}  \frac{1+\cos^2\tilde{\theta}_L}{2}\cos[\omega  (t - \tilde{r})-2\tilde{\phi}_L]\,,\\
\label{CrossRealExpression}
h_\times(t, \tilde{r}, \tilde{\theta}_L, \tilde{\phi}_L) &= \frac{A_\text{in}}{\tilde{r}} \cos \tilde{\theta}_L\sin[\omega (t - \tilde{r})-2\tilde{\phi}_L]\,,
\end{align}
where
\be\label{calA}
A_\text{in} = 4 M_c^{5/3} (\pi f)^{2/3}\,,\qquad \omega = 2 \pi f\,,
\ee

which holds along the source-lens direction  $(\tilde{\theta}_L, \tilde{\phi}_L)$, i.e. $\tilde{r}$ denotes the radial distance from the source in the fixed $e_z$ direction. Note that the dependence on the phase $\tilde{\phi}_L$ corresponds to a physically arbitrary choice of initial phase in the binary orbital motion, which propagates in the relative choice of orientation for $\tilde{e}_{x,y}$ in Fig.~\ref{MovingScheme}. 
Rather than reabsorbing it, we keep track of this phase anticipating the treatment of a superposition of scattered and transmitted waves.

We are interested in considering the wave impacting on the lens, i.e. we rewrite the radial variable as $\tilde{r} = d_\text{SL}+z$, with $d_\text{SL}$ distance source-lens, and $z$ the line of sight distance measured from the lens (along the source-lens direction).
In this $(\tilde{\theta}_L, \tilde{\phi}_L)$ direction, one can check that the local plane wave can be written as\footnote{Note that $(e_x, e_y)$ of the source and the lens coincide, hence expressions below can be interpreted as a wave expansion around the lens.} \\
$h_{ij}\propto$
\begin{align}
e^{\ii(k\tilde{r}-\omega t)}\,e^{\ii 2\tilde{\phi}_L} &\left(
\begin{array}{ccc}
\frac{1+\cos^2\tilde{\theta}_L}{2}&\ii\cos\tilde{\theta}_L&0\\
\ii\cos\tilde{\theta}_L&-\frac{1+\cos^2\tilde{\theta}_L}{2}&0\\
0&0&0\\
\end{array}
\right)_{ij}\!\! \cc\,\\
\propto \  e^{\ii(k\tilde{r}-\omega t)} & \left\lbrace \,{}_2Y_{22}(\tilde{\theta}_L, \tilde{\phi}_L)\left(
\begin{array}{ccc}
1&-\ii&0\\
-\ii&-1&0\\
0&0&0\\
\end{array} \right)_{ij} \right.\nonumber\\& 
\left. \,{}_{-2}Y_{22}(\tilde{\theta}_L, \tilde{\phi}_L)\left(
\begin{array}{ccc}
1&\ii&0\\
\ii&-1&0\\
0&0&0\\
\end{array} \right)_{ij}  \right\rbrace \cc
\end{align}
For a locally plane wave in the lens neighbourhood,
\be
\frac{e^{\ii(k \tilde{r}-\omega t)}}{\tilde{r}} = \frac{e^{\ii (k d_\text{SL}-\omega t)}}{d_\text{SL}}\,\frac{e^{\ii k z}}{1+\frac{z}{d_\text{SL}}}\simeq \frac{e^{\ii (k d_\text{SL}-\omega t)}}{d_\text{SL}}\, e^{\ii k z}\,,
\ee
we have
\be\label{H}
 h_{ij}=\sum_{m=\pm 2} H^{(m)} Q^{2m}_{ij} \cc \,,
\ee
with
\begin{equation}
Q^{2\, \pm 2}_{ij} = -\sqrt{\frac{3}{8}}{(e_x \pm \ii e_y)}_i \ {(e_x \pm \ii e_y)}_j {\rm e}^{\ii k z}\,,
\end{equation}
and 
\begin{align}
&H^{(m=\pm2)} =\nn\\
&\frac{-2}{d_\text{SL}} \sqrt{\frac{2\pi}{15}}A_\text{in}\ {}_{\mp2}Y^{22}(\tilde{\theta}_L,\tilde{\phi}_L)\ e^{\ii \omega ( d_\text{SL} - t)}\,.
\label{hpm2factors}
\end{align}
The choice of normalisation for the basis $Q_{ij}^{2m}$ in Eq.~(\ref{H}) is taken in agreement with Refs.~\cite{Hu:1997hp} and \cite{Pitrou:2019ifq}, allowing to use results therein for the decomposition of a plane wave (with spin) onto a basis of spin-weighted spherical harmonics. 
We observe that the  modes $H^{(m)}$ can be related to standard helicity modes $h^{(\pm 2)}$, defined as
\begin{align}
h^{(\pm2)}&=h_+ \pm \ii h_{\times}\,.
\end{align}
The relation is given by\footnote{This is derived writing the wave as $h_{ij}=\Theta_{ij}^{(2m)}h^{(m)}$ with $\Theta^{(2\pm2)}=1/2(e_x\pm \ii e_y)_i(e_x\pm \ii e_y)_j$ and comparing with (\ref{H}).}  
\begin{align}
h^{(-2)}&=-\sqrt{\frac{3}{2}}\left(H^{(2)} e^{\ii k z} +H^{(-2)*} e^{-\ii k z} \right)\,,\\
h^{(+2)}&=-\sqrt{\frac{3}{2}}\left(H^{(-2)} e^{\ii k z} +H^{(2)*} e^{-\ii k z} \right)\,.
\end{align}

\subsection{Final form of the incoming wave}

  We want to identify the even and odd Regge-Wheeler potentials associated with the plane wave \eqref{H}, in a system of spherical coordinates centered on the lens. To do so, we need to decompose the plane wave in spherical components (radial and tangential) in spherical coordinates, i.e. find the explicit expression for the functions appearing in the decomposition (\ref{pAB})-(\ref{pABf}). Details can be found in appendix \ref{matching}. Given the explicit expression of the plane wave, the Regge-Wheeler variables (\ref{Psioddfromhr}) and (\ref{Psioddfromhr_even}) associated to it are found by using the corresponding functional form of the radial functions $h^{\ell m}_{rr},K^{\ell m},G^{\ell m},h^{\ell m}_2,h^{\ell m}_r,j^{\ell m}_r$ (other components vanish).

 We obtain that the complexified incoming wave (emitted by the binary) around the lens, in the lens frame corresponds to the Regge-Wheeler variables
    \begin{widetext}
\begin{align}
    &\Psi^{\ell m}_\text{even} =\ii\frac{H^{(m)}}{2k}\,\sqrt{6\pi}\sqrt{2\ell+1}\sqrt{\frac{(\ell-2)!}{(\ell+2)!}} \, \left[(-1)^{\ell+1} e^{- \ii k r}+e^{\ii k r}\right] \cm\,,\label{PsiEvenNoScattForm} \\
&\Psi^{\ell m}_\text{odd} =- \frac{m}{|m|}\frac{H^{(m)}}{2k}\,\sqrt{6\pi}\sqrt{2\ell+1}\sqrt{\frac{(\ell-2)!}{(\ell+2)!}} \, \left[(-1)^{\ell+1} e^{- \ii k r}+e^{\ii k r}\right] \cm \,,
\label{PsiOddNoScattForm}
\end{align}

where c.m. refers to the ``conjugate mode'': for a given $\Psi_\bullet^{\ell, m}$, its conjugate mode is $(-1)^m \Psi_\bullet^{\ell, -m*}$ as introduced in Appendix~\ref{matching}. 
Eqs.~(\ref{PsiEvenNoScattForm}) and (\ref{PsiOddNoScattForm}) are the main result of this section.
\end{widetext}

\section{Scattered wave}\label{sec_scatt}

We compute the scattered wave. Indeed, we know that Regge-Wheeler variables obey the scattering equation~\eqref{ReggeWheeler} and that, after scattering, they are related to the ones before scattering via a phase shift; see Eqs.~(\ref{PhaseShiftMinus}) and (\ref{dd1}).  
The radiated polarisations at infinity are recovered from Ref.~\cite{Martel:2005ir}

\be
h^{(\pm 2)}=\sum_{\ell m} \,_{\pm2}Y^{\ell m}h^{(\pm 2)}_{\ell m}\,,
\label{hpmReconstructionSpherical}
\ee
and 
\be\label{Eq_hlm2}
h^{(\pm 2)}_{\ell m}=\frac{1}{2r}\left[   \frac{(\ell+2)!}{(\ell-2)!}\right]^{1/2}
( \Psi^{\ell m}_\text{even}\pm \ii  \Psi^{\ell m}_\text{odd}) \,.
\ee
Note that for $h^{(\pm 2)}_{\ell m}$ multipoles indices $\ell m$ are put down rather than up for ease of reading, in practice we use both up/down notation interchangeably. 

Here polarisations $h_{+, \times}$ are defined with respect to the basis $(e_\theta, e_\phi)$ in the polarisation plane transverse to the propagation direction $e_r$.

To extract the scattered part of the wave which we are interested in, we use the scattered part of the $\Psi^{\ell m}_\text{even, odd}$ master functions in Eq.~\eqref{Eq_hlm2}. Following the procedure outlined in Sec.~\ref{sec:PhaseShift}, the scattering process modifies the full $\Psi^{\ell m}_\text{even, odd}$ waves by altering the spherically outgoing partial waves with the phase shifts given by \eqref{PhaseShiftMinus} and \eqref{dd1}. The scattered part then consists of these full waves to which we remove all the contributions from the initial incoming wave. The coordinate $r$ is further promoted to $r_*$ in the phase of the functions, and accordingly for the incident radial quantity $d_\text{SL}$, since we use phase shifts computed with respect to this radially corrected phase \cite{PoissonSasaki}.  We find that the scattered modes are given by 
\begin{widetext}
\begin{align}
    \label{hlm}
&h_{\ell m}^{(\pm 2)} = \ii \frac{\sqrt{6\pi(2\ell+1)}}{4kr} \  \nn \\ \Bigg( &
H^{(m)} \left[(e^{2 \ii \delta_{\ell}^{\text{even}}}-1) \mp(e^{2 \ii \delta_{\ell}^{\text{odd}}}-1)\frac{m}{|m|} \right] e^{\ii kr_*} - (-1)^m H^{(-m)*} \left[(e^{-2 \ii \delta_{\ell}^{\text{even}}}-1) \mp(e^{-2 \ii \delta_{\ell}^{\text{odd}}}-1)\frac{m}{|m|} \right] e^{-\ii kr_*}\Bigg).
\end{align}    
\end{widetext}
This expression has to be inserted in Eq.~(\ref{hpmReconstructionSpherical}) to compute the resummed series defining helicity modes and $+$ and $\times$ polarisations. A detailed discussion of the convergence and summability of this series can be found in the dedicated appendix~\ref{sec:DivergingSeries} while its resummation is described in appendix \ref{app:scattered}. The final asymptotical result for the scattered polarisations is 
\begin{widetext}
\begin{align}
h_+ &= \  \frac{2M}{r} \frac{A_\text{in}}{d_\text{SL}}\frac{1}{(1-\cos\theta)} \frac{1+\cos^2\theta}{2}{}\left[\cos^4\left(\frac{\tilde{\theta}_L}{2}\right)\cos\left(\varphi-2\phi\right)+\sin^4\left(\frac{\tilde{\theta}_L}{2}\right)\cos\left(\varphi+2\phi\right)\right]\,,\label{hplScattered}\\
h_\times &= \frac{2M}{r} \frac{A_\text{in}}{d_\text{SL}}\frac{1}{(1-\cos\theta)} \cos(\theta) \left[\cos^4\left(\frac{\tilde{\theta}_L}{2}\right)\sin\left(\varphi-2\phi\right)-\sin^4\left(\frac{\tilde{\theta}_L}{2}\right)\sin\left(\varphi+2\phi\right)\right]\,,\label{hxScattered}
\end{align}
\end{widetext}

with $A_\text{in}$ is defined in (\ref{calA}), and a phase $\varphi(r, t)$ defined as 
\be\label{eq.phaselog}
\begin{split}
\varphi(r, t) \equiv \ &\omega(t -d_{\text{SL}*}-r_*) - 2\tilde{\phi}_L \\ +&\Phi-  2M\omega\left[\ln\left(1-\cos\theta\right)-1-\ln 2\right]
\,,
\end{split}
\ee
where we recall that $\Phi$ is defined after Eq.~(\ref{PhaseShiftMinus}).  Equations~(\ref{hplScattered}) and (\ref{hxScattered}) describing the scattered wave are the main results of this section. They are expressed for any $(r, \theta, \phi)$ observer location, in the polarisation basis $ (e_\theta, e_\phi)|_\text{O}$. These results have the structure of a double projection of the original source quadrupole, as is further detailed in Appendix~\ref{app:scattered}.

The $\big(1 - \cos\theta\big)^{-1}$ divergence in the limit $\theta\to0$ is expected from the corresponding Coulomb problem, i.e. Rutherford scattering, which can be treated in multipole space analogously to our problem; see Ref.~\cite{Rutherford2022} for a detailed study that includes the analogy with scalar waves on Schwarzschild spacetime. In the Coulomb case, an exact solution exists, of which the $\big(1 - \cos\theta\big)^{-1}$ expression is only an approximation in the asymptotically limit of $\omega r \big(1 - \cos\theta\big)\gg1$, thereby explicitly excluding $\theta=0$. Since the Rutherford scattering is free from any divergence issue (and a similar feature is expected for scalar wave propagation on Schwarzschild spacetime~\cite{Rutherford2022}), the same may naturally be expected for GW scattering, despite the fact that in this context there is no known exact closed form solution. 
One may thus use the regime of validity of the latter, i.e. $\omega r \big(1 - \cos\theta\big)\gg1$ as a benchmark for the region of validity of our GW plane wave scattering process derivation.

Computing the two helicity modes $h^{\pm2}$ from Eqs.~(\ref{hplScattered}) and (\ref{hxScattered}), and comparing with the incident helicity, one can explicitly check that the scattered and incident helicity are in general different. The (absolute value of the) ratio of the two helicity modes is not conserved during propagation on Schwarzschild spacetime, unlike what happens in geometric optics.  This fact is rooted in the scattering process being not helicity preserving \cite{deLogi,Dolan:2007ut}. The appropriate quantity to illustrate this property is the degree of circular polarisation $\mathcal{V}$ of a wave, that will be introduced in Sec.\ref{sec_cross}. 

\section{Cross section}\label{sec_cross}

We now consider the differential cross section of the scattering process. We recall that in an asymptotically flat universe, a GW carries an energy flux, that, for a transverse-tranceless $h_{ij}$ reads \cite{Maggiore}
\be
\frac{\dd E}{\dd A \dd t}=\frac{1}{32 \pi } \langle \dot{h}_{ij}\dot{h}_{ij}\rangle\,,
\ee
where $\dd A=r^2 \dd\Omega$ is the area element at radial distance $r$, and $\langle \,\cdots\rangle$ denotes time averaging over a few periods. The differential scattering cross section is then defined as the ratio of outgoing differential (i.e. angular) energy flux to the incident energy flux, that is
\be\label{domega}
\frac{\dd\sigma}{\dd\Omega} = \frac{\dd E}{\dd\Omega \dd t}\bigg\vert_\text{scat} \bigg/  \frac{\dd E}{\dd A \dd t}\bigg\vert_\text{in} = r^2 \frac{\langle \dot{h}_{ij}\dot{h}_{ij}\rangle\vert_\text{scat}}{\langle \dot{h}_{ij}\dot{h}_{ij}\rangle\vert_\text{in}}\,.
\ee
From Eqs.~\eqref{hplScattered} and \eqref{hxScattered}, the numerator of  Eq.~(\ref{domega}) reads
\begin{align}
r^2 \langle \dot{h}_{ij}\dot{h}_{ij}\rangle\vert_\text{scat} = &\left(\frac{A_\text{in}}{d_\text{SL}}\right)^2 \omega^2M^2\Bigg[ \frac{1}{4}\sin^4\tilde{\theta}_L \cos^4\left(\frac{\theta}{2}\right) \cos(4 \phi)\nn\\ &+ \Bigg(\cos^8\left(\frac{\theta}{2}\right)+\sin^8\left(\frac{\theta}{2}\right)\Bigg)  \sin^{-4}\left(\frac{\theta}{2}\right) \nn\\&\times\Bigg(\cos^8\Bigg(\frac{\tilde{\theta}_L}{2}\Bigg)+\sin^8\Bigg(\frac{\tilde{\theta}_L}{2}\Bigg)\Bigg) \Bigg]\,.
\end{align}

For the incoming wave, we have 
\be
\langle \dot{h}_{ij}\dot{h}_{ij}\rangle\vert_\text{in}  =\left(\frac{A_\text{in}}{d_\text{SL}}\right)^2 \omega^2 \left(\cos^8\left(\frac{\tilde{\theta}_L}{2}\right)+\sin^8\left(\frac{\tilde{\theta}_L}{2}\right)\right)\,.
\ee
Therefore, the differential cross section can be written as
\begin{align}
\frac{\dd\sigma}{\dd\Omega} =& M^2 \frac{\cos^8\left(\frac{\theta}{2}\right)+\sin^8\left(\frac{\theta}{2}\right)}{\sin^4\left(\frac{\theta}{2}\right)}\quad \nonumber\\
+& \frac{1}{4}M^2 \frac{ \sin^4\tilde{\theta}_L}{\sin^8 \left(\frac{\tilde{\theta}_L}{2}\right)+\cos^8\left(
   \frac{\tilde{\theta}_L}{2}\right)} \cos^4\left(\frac{\theta }{2}\right) \cos (4 \phi )\,.
   \label{FulldsigmadOmega}
\end{align}
This result agrees with what is obtained in the literature for special geometrical configurations, see appendix  \ref{app_cross}. The $\left[\sin(\theta/2)\right]^{-4}$ behaviour when  $\theta\to0$ is a shared feature with Rutherford scattering, as previously mentioned.

We further observe that the first contribution to the differential cross section depends only on the geometry between the lens and the observer. Contrariwise, the second contribution depends both on the geometry between source and lens, and on that between lens and observer. However, this second contribution  vanishes for exactly two source-lens configurations, namely $\tilde{\theta}_L = 0, \pi$. These particular configurations correspond exactly to those for which the incoming plane wave has $+$ and $\times$ polarisations of the same (absolute) amplitude, as is easily seen from Eqs.~\eqref{PlusRealExpression}, \eqref{CrossRealExpression}.

For all other values of $\tilde{\theta}_L$, the incoming plane wave has a preferred polarisation by virtue of the geometry between the source and the lens. The resulting scattering differential cross section carries the imprint of this preferred polarisation in its second term, while the first term is polarisation independent.

Based on that, we investigate how this $\tilde{\theta}_L$-dependence is related to the degree of circular polarisation of the incoming plane wave. We introduce the latter from the Fourier-modes $\tilde{h}_{+, \times}(f)$ of the polarisations: 
\be\label{V}
\mathcal{V} = \frac{2\text{Im}[\tilde{h}_+\tilde{h}_\times^*]}{|\tilde{h}_+|^2+|\tilde{h}_\times|^2}=\frac{|\tilde{h}^{(2)}|^2-|\tilde{h}^{(-2)}|^2}{|\tilde{h}^{(2)}|^2+|\tilde{h}^{(-2)}|^2}\,,
\ee
where it is understood that one must consider only the prefactors to the Dirac distributions (the latter being inherently present due to the assumed monochromatic signal).

Note that $\mathcal{V}$  corresponds to the ratio of the Stokes parameter $V$ describing circular polarisation and the intensity $I$, see Ref.~\cite{Cusin:2018rsq}. $\mathcal{V}$ is bounded between -1 (for pure helicity -2) and 1 (for pure helicity 2), and vanishes for linearly polarised waves. It is further invariant under a rotation of the basis used in the plane transverse to the direction of propagation of the wave. One can verify that the polarization dependent part of the cross section~(\ref{V}) can be re-expressed in terms of the degree of polarization of the incoming wave, using that 
\be
\frac{ \sin^4\tilde{\theta}_L}{\sin ^8\left(\frac{\tilde{\theta}_L}{2}\right)+\cos^8\left(
   \frac{\tilde{\theta}_L}{2}\right)} = 8\sqrt{1-\mathcal{V}^2_\text{incident}}\,,
\ee
where $\mathcal{V}_\text{incident}$ is computed from the incident $\tilde{h}_{+\times}$ at the lens.

Consider a situation in which the degree of circular polarisation of the impinging wave is exactly~\,0 ($\tilde{\theta}_L=\pi/2$).
Then, an observer at $\theta = \pi/2$  would be placed at a location where the scattering cross section is $\frac{1}{2}M^2 \big[1+\cos(4\phi)\big] $, by Eq.~\eqref{FulldsigmadOmega}. In that particular configuration, depending on the observer's azimuthal position $\phi$ with respect to the lens, the cross section may be doubled (when $\phi = n\pi/2, n\in \mathbb{N}$) with respect to that of a circularly polarised wave, or may vanish (when $\phi = (1+2n)\pi/4, n\in \mathbb{N}$).

\begin{figure}
\includegraphics[width = 0.9\columnwidth]{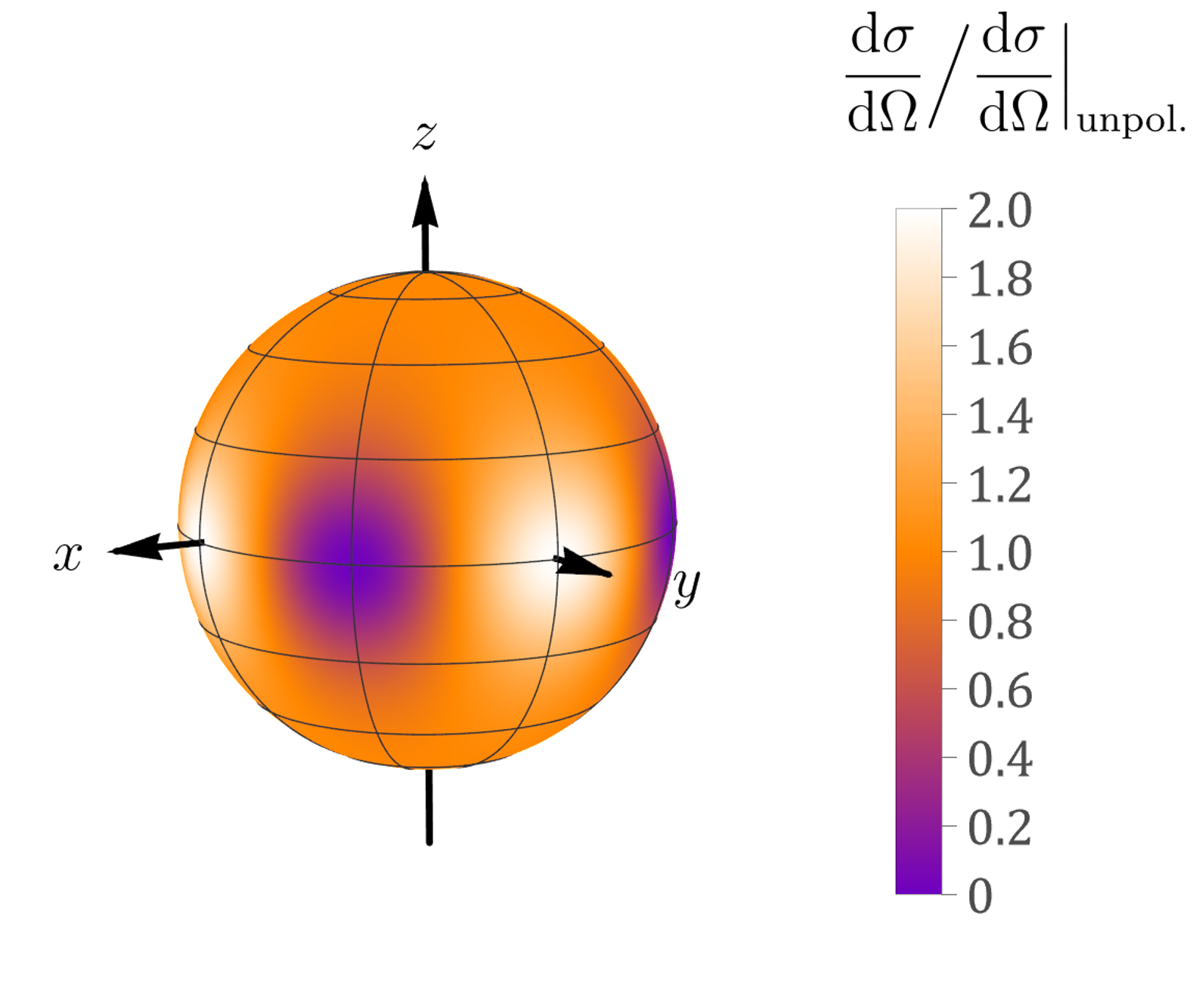}
\caption{Ratio of the total cross section to the polarisation independent-part of the cross section as a function of observer angular position, for a purely $h_+$ incoming plane wave from the $-z$ direction reaching a lens situated at the origin. This corresponds to fixed $\tilde{\theta}_L= \pi/2$ so that the incident wave has no circular polarisation.
While for forward (positive $z$) and backward (negative $z$) scattering, the total cross section is dominated by the polarisation independent cross section, the polarisation dependent part is significant for scattering angles $\theta$ about $\pi/2$.}
\label{fig:CrossSectionRatio}
\vspace{-0.5cm}
\end{figure}

This is illustrated in Fig~\ref{fig:CrossSectionRatio}, where we display the ratio of the total cross section to the polarisation independent-part of the cross section (i.e. the first term of Eq.~\eqref{FulldsigmadOmega}) for the particular source-lens geometry mentioned here.

\medbreak
We have mentioned the condition $\omega r (1-\cos\theta)\gg1$ as a benchmark for the regime of validity of our results. It may be used to define a minimal value for $\theta = \theta_\text{min}$ where the results may be used, given $\omega r$, for a plane GW scattering process. While we effectively solve a plane wave scattering equation, we should recall that the real case of interest involves the scattering of a spherical wave that is only locally plane. A spherical wave is perceived as locally plane if $d_\text{SL}$, the distance from the emitter, is much larger than the characteristic size of the scattering center. More than the lens's Schwarzschild radius $2M$, the (integrated) cross section naturally provides a constraining effective length scale that is relevant for the scattering process. In this view, on top of the validity condition $\omega r (1-\cos\theta)\gg1$ of the plane wave computation, we may set the further constraint that for spherical waves, the results should remain valid for values of $\theta_\text{min}$ such that the (incompletely) integrated cross section 
 \be
  \int_{\theta_\text{min}}^\pi \
 \dd\theta \int_{0}^{2\pi} \dd\phi \sin \theta \
 \frac{\dd\sigma}{\dd\Omega} < 10\%\ \pi d_\text{SL}^2\,,
 \label{ValidityCriterion}
 \ee
 where 10\% is an arbitrary threshold.
 Below this $\theta_\text{min}$ value, the scattering process associates an effective size to the scattering center that is too large compared to $d_\text{SL}$ for the incident wave to be judged locally plane.
 For cases of interest, the criterion of local planarity \eqref{ValidityCriterion} is more constraining than the regime of validity of plane wave scattering $\omega r (1-\cos\theta_\text{min})\gg1$. We observe that such criteria would in principle not be necessary if an analytic framework designed to study scattering of spherical waves existed for our context. 
 
\section{Total wave}\label{sec_totalwave}

In standard scattering theory, the total wave received by the observer is given by the superposition of scattered and transmitted wave. These two waves will interfere, creating an interference pattern in the observed total waveform. The scattered wave is given by Eqs.~(\ref{hplScattered}) and (\ref{hxScattered}), however the computation of the transmitted part is not trivial.

In our case the source emits a non isotropic spherical wave that we have approximated by a plane wave only at the lens scale. 
It is clear that as $M\to 0$, the transmitted wave should reduce to the incident spherical wave evaluated at the observer position, and assuming that this is the correct answer also for   $M\neq0$ has been proven valid in electromagnetic and acoustic scattering processes involving spherical waves interacting with a sphere \cite{Bowman}, and this is what we assume in our context for illustration\footnote{The validity of this assertion in our case should be assessed in a rigorous framework of gravitational spherical wave scattering.}. In other words, we assume that the transmitted wave is what we would find if we computed the two polarisation models out of $\Psi^{\ell, m}_\text{even, odd}$ without adding the scattering phase shifts, reversing the operations of section \ref{sec_in}.

We observe that when computing the total wave in the analogous context of scattering of scalar wave off Coulomb potentials, the incident wave has to be corrected by an angle dependent logarithmic phase modification for which there exists a closed form, see e.g. \cite{Rutherford2022}. 

A closed result for the phase shift in the case of a spherical wavefront does not exist to date. 
Nevertheless, any non trivial phase correction would only modify the resulting interference pattern between incoming and scattered wave, but not the order of magnitude of the interference effect, i.e. the strength of scattering phenomenon.\footnote{A quantitative prediction of the total wave would require to numerically solve the scattering problem (probably making use of algorithms from numerical relativity), and including line of sight effects that have been neglected in our study. We postpone this to a future work.}

For illustrative purposes, in section \ref{sec_pheno}, we will thus approximate the transmitted wave as the incident spherical wave emitted by the source in the observer (O) direction, i.e. \eqref{PlusRealExpression}, \eqref{CrossRealExpression} evaluated at observer rather than lens position:
\begin{align}
\label{TransPlusExpression}
h_+^\text{trans} &= \frac{A_\text{in}}{d_\text{SO}} \frac{1+\cos^2\tilde{\theta}_\text{O}}{2}\cos[\omega( t - d_{\text{SO}})-2\tilde{\phi}_\text{O}]\,,\\
\label{TransCrossExpression}
h_\times^\text{trans}  &= \frac{A_\text{in}}{d_\text{SO}} \cos\tilde{\theta}_\text{O}\sin[\omega (t - d_{\text{SO}})-2\tilde{\phi}_\text{O}]\,,
\end{align}
 acknowledging that the exact expression will have additional phase corrections. 

When summing the scattered and transmitted waves to build the total wave at observer, we should make sure to make a unique choice of polarisation basis with respect to which $h_{+, \times}$ are expressed. Indeed, the scattered $h_{+, \times}$ \eqref{hplScattered}, \eqref{hxScattered} are expressed in a polarisation basis $(e_\theta, e_\phi)|_\text{O}$, transverse to the propagation direction $e_r|_\text{O}$, while the aforementioned transmitted polarisations would be expressed in a polarisation basis $(\tilde{e}_{\tilde{\theta}}, \tilde{e}_{\tilde{\phi}})|_\text{O}$, transverse to the propagation direction $\tilde{e}_{\tilde{r}}|_\text{O}$. In the distant observer approximation, the two propagation directions match, $\tilde{e}_{\tilde{r}}|_\text{O} \simeq e_r|_\text{O}$, but transverse to that, the polarisation modes should be properly rotated to be added. We choose to rotate the polarisation basis associated to the scattered wave into that of the transmitted wave. 
That is, for scattered $h_{+, \times}$,
\begin{align}
h_+ &\to h_+ \cos(2\eta) - h_\times \sin(2\eta)\,,\\
h_\times &\to h_+ \sin(2\eta) + h_\times \cos(2\eta)\,,
\end{align}
with $\cos\eta = (\tilde{e}_{\tilde{\theta}}\cdot e_\theta)|_\text{O}  = (\tilde{e}_{\tilde{\phi}}\cdot e_\phi)|_\text{O}$ and $\sin(-\eta) = (\tilde{e}_{\tilde{\phi}}\cdot e_\theta)|_\text{O}$.

\section{Phenomenological implications}\label{sec_pheno}

Having at hand analytical expressions for the quantities involved in a process of wave optics lensing, we now attempt to derive phenomenology.
To quantify the strength of the scattering effect, we note that leaving aside all angular geometrical factors but the scaling at small angles, the amplitude ratio $\rho$ of scattered to transmitted wave is [e.g. compare Eqs.~\eqref{hplScattered} and \eqref{TransPlusExpression}]
\be\label{relmod}
\rho \equiv 2M\frac{d_\text{SO}}{d_\text{SL}d_\text{LO}}\frac{1}{(1-\cos\theta)} \simeq \frac{2M}{d_\text{SL}}\frac{1}{(1-\cos\theta)}\,.
\ee
We identify two important factors governing the amplitude of the scattered wave in terms of the transmitted one: the source-lens distance, in units of the lens Schwarzschild radii and the alignment of the lensing geometry.  The work  by Ref.~\cite{Bellovary:2015ifg} shows that stable orbits around an AGN are found in disc migration traps located at $d_\text{SL} \sim 24.5 \times 2M$ and $d_\text{SL} \sim 331\times 2 M$ (assuming a particular disc model). For $d_\text{SL} \sim 50 M$ one expects orbital velocities of order $\mathcal{O}(0.1)$ and there should be relativistic orbital dynamics involved. 
For the sake of ease in illustration, we thus consider $d_\text{SL}= 331\times 2M$.

Then, the scattered wave is suppressed with respect to  the transmitted wave by a factor of $10^{-3} (1-\cos\theta)^{-1}$, where the factor $(1-\cos\theta)^{-1}$ can reach $\sim 80$ for e.g. $\theta\sim0.15$, which is still within the regime of validity given by see Eq.~\eqref{ValidityCriterion}. 
In addition to this forward magnification, to get a large relative modulation effect, one can also think of geometrical configurations where one of the transmitted polarisations is trigonometrically strongly suppressed while the corresponding scattered polarisation is not. In such a case the scattered to transmitted amplitude ratio may excess 50\%, leaving a strong imprint in the total wavefront.

To get a detectable modulation signal, one has to make sure that the transmitted wave (i.e. what one would get without lensing) is detectable with a sufficiently high SNR. This is the case for systems hosted in the low redshift tail of the AGN distribution, e.g. $z\sim0.01$ \cite{CAIXAcatalogueAGN}. Using \cite{RobsonLISA} to estimate SNR for LISA and adapting results of \cite{Sberna:2022qbn}, we find that a  hierarchical triple system like ours with $M_c\sim 80 M_\odot$, $f\sim 0.03$Hz is detectable by LISA with SNR
above 100 for a 4-year mission, if located at $z\sim0.01$ \footnote{It was found by \cite{Sberna:2022qbn} that such a hierarchical triple is detectable by LISA with SNR of $5-10$ for a 4-year mission, if located at $z\simeq 0.27$. Transposing this system to a lower $z\sim0.01$ would allow a boost in SNR for the transmitted wave by a factor $\sim d_\text{SO}(z = 0.27)/d_\text{SO}(z =0.01) \sim 30$.}.
This implies that the modulation in the total wave due to interference between scattered and transmitted wave would be detectable for such a signal, in the most favourable geometrical configurations.

In this phenomenological illustration, we stick to the reference case of a $M_c= 80M_\odot$ inspiral of Ref.~\cite{Sberna:2022qbn} and consider the optimistic scenario where the lens is located at the low end of the AGN redshift distribution, i.e. $z\sim0.01$ \cite{CAIXAcatalogueAGN}, with $d_\text{SL} = 662 M$. Note that with the choice $M_c= 80M_\odot$, the system has  massive components reaching the intermediate mass gap. However, AGN migrations traps have been found to be a favorable environment for the production of black holes in this mass range \cite{Bellovary:2015ifg}, making the assumption reasonable. 
We emphasise that any choice of source redshift and chirp mass gives only a scaling of the overall amplitude of the total wave (and thereby the SNR), but does not impact the relative strength of the scattering effect that we are describing, which is fully given by the geometry once $d_\text{SL}$ is fixed. 
As for the $M\omega\ll1$ requirement, we take $\omega = 2\pi f=2\pi\times 3\times 10^{-3}$Hz to be within the LISA band, and consider an AGN mass of $M = 1.2\times 10^6 M_\odot$ to reach $M\omega\simeq0.11$. Such an AGN mass is not unrealistic even at $z\sim10^{-2}$, as is shown in \cite{NGC4051}\footnote{We observe that for a binary with equal mass and the geometry chosen above, their separation is $2.2\times 10^5 R_\odot$, hence very far from merging but very small compared to $d_\text{SL}$.}. 

So far, we considered a static lensing configuration.  
In realistic situations, the binary of solar mass objects orbits around the central black hole, as was depicted in Fig.\,\ref{MovingScheme}. To study this dynamical setting, we evaluate our static scattering results letting geometrical quantities become functions of time (in other words, we do not include any further kinematic effect apart from the evolution of the geometry and the corresponding time-delay).

To study the evolution of the waveform with time, we need to parameterise the motion of the binary around the AGN, and to take into account how the various angular quantities in Eqs.~(\ref{hplScattered}) and (\ref{hxScattered})  evolve with time. For illustration, and motivated by the stability of the disk migration traps, we assume that the source is in a Newtonian circular orbit around the AGN, under the approximation of axial parallelism (no precession) for the inspiraling binary system. We do not include any relativistic effect in the motion. Our description of the dynamics is presented in detail in appendix \ref{triple}. There, three relevant free parameters that specify the scattering geometry, are introduced: $(\iota, \alpha_{1, 2})$. The angle $\iota$ describes the inclination of the orbital plane of the outer binary with respect to the source-lens axis (see Fig.\,\ref{MovingScheme}). It corresponds to the minimal scattering angle $\theta$ that is reached along the orbit. Note that by fixing  $\iota\neq0$, any forward divergence is avoided. When we consider a time varying scattering geometry, the source is sending a varying polarisation to the lens over time, because the emitted signal is not isotropically polarised. The angles  
$\alpha_{1, 2}$ describe the orientation of source's angular momentum with respect to the observer-lens rest frame, and thereby parametrise the emitted polarisation.

Under the assumption of circular motion, and for our mass and distance values the orbital period is 
\be
T=\sqrt{\frac{4\pi^2 d_\text{SL}^3}{M}}\simeq 1 {\rm \  week}\,,
\ee
hence for such a system one expects to observe multiple periods of revolution of the binary around the AGN. The scattering effect is thus periodic as long as the binary emitting frequency and distance to the third body remain constant. 
Using $\Omega = 2\pi/T$ and Eq.~\eqref{relmod}, the relative strength $\rho$ of the effect has the maximal value
\be
\label{relmod2}
\rho_\text{max} = 2\left(M\Omega\right)^{2/3}\, (1-\cos\iota)^{-1}\,.
\ee

\begin{figure}[b]
\includegraphics[width = \columnwidth]{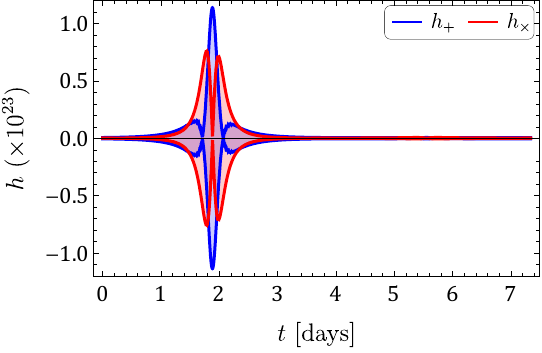}
\caption{Configuration $\iota = 0.15,\, \alpha_1 = 0.16,\, \alpha_2 = 6.12$. Envelope of scattered waves \eqref{hplScattered} and \eqref{hxScattered}, for both polarisations, over the course of one orbit. The strong $(1-\cos\theta)^{-1}$ amplification is taking place at the point of best alignment. Even in the absence of the $(1-\cos\theta)^{-1}$ amplification, for this configuration the scattered waves show strong interference patterns between the two subcomponents of \eqref{hplScattered}, resp. \eqref{hxScattered}. These submodulations of the polarisation's amplitudes are correspondingly amplified at small inclination. }\label{ScatteredWaves}
\end{figure}

Figure \ref{ScatteredWaves} shows the scattered wave polarisations given by Eqs.~\eqref{hplScattered} and \eqref{hxScattered} over the course of one orbit, for a particular configuration. 
The inclination $\iota$ is chosen so as to have sizeable amplification factor, while remaining in the bound of Eq.~\eqref{ValidityCriterion}. We see that the amplitude of both polarisation modes increases at a certain point in the orbit, which coincides with the situation of best source-lens-observer alignment.
It is interesting to look at the total wave, sum of transmitted and scattered components, which is the observable quantity at the observer. 

\begin{figure}[b]
  \includegraphics[trim=0 0.1mm 0 0, clip,width =\columnwidth]{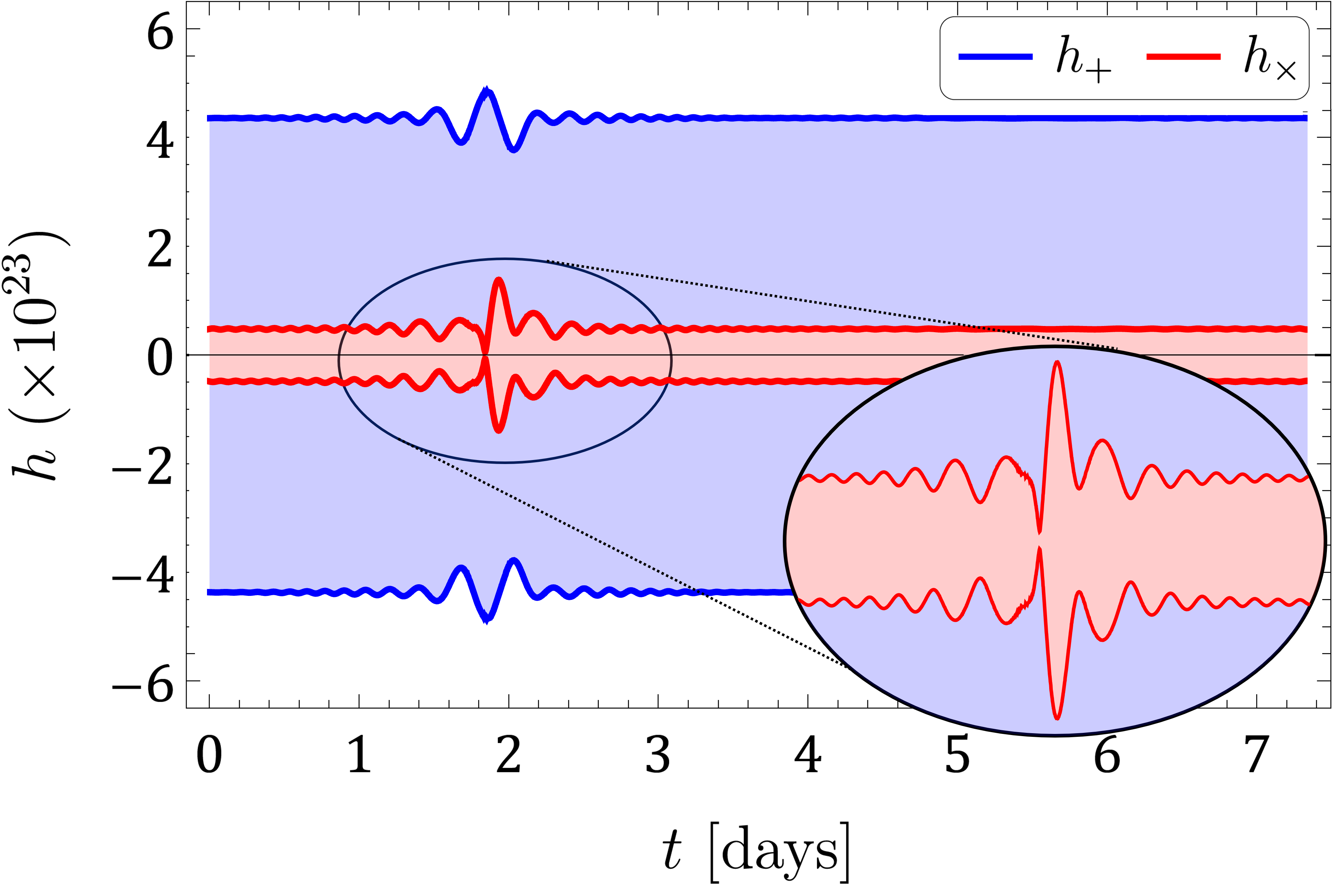}\\
  \includegraphics[width =\columnwidth]{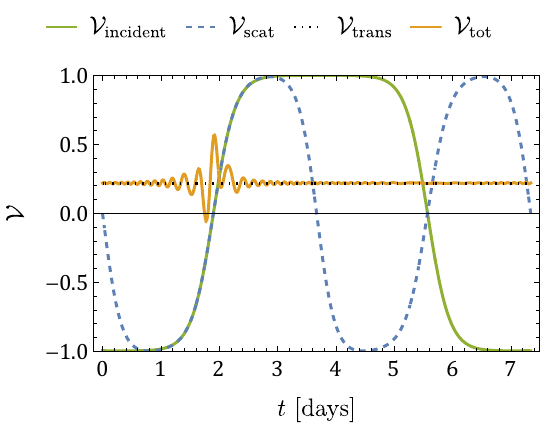}
\caption{Configuration $\iota = 0.16,\, \alpha_1 = 0.90,\, \alpha_2 = 1.50$ \\ \textit{Top }: Envelope of the total wave, for both polarisations, over the course of one orbit in a configuration of low total circular polarisation. For this configuration with $h_+ > h_\times$, the relative modulation due to interference is of order 12\% on $h_+$ and excesses 180\% on the (weak) $h_\times$ signal. 
 \textit{Bottom} : Corresponding degree of circular polarisation. We show $\mathcal{V}_\text{incident}$ as the degree of circular polarisation of the incoming wave at the lens, $\mathcal{V}_\text{scat}$ that of the above scattered wave, $\mathcal{V}_\text{trans}$ the (constant) one of the transmitted wave at the observer, and $\mathcal{V}_\text{tot}$ that of the total wave. The oscillations in the individual polarisations amplitudes propagate in the total degree of circular polarisation.}\label{TotalWavesLowV}
 \end{figure}

\begin{figure}[b]
\includegraphics[width=\columnwidth]{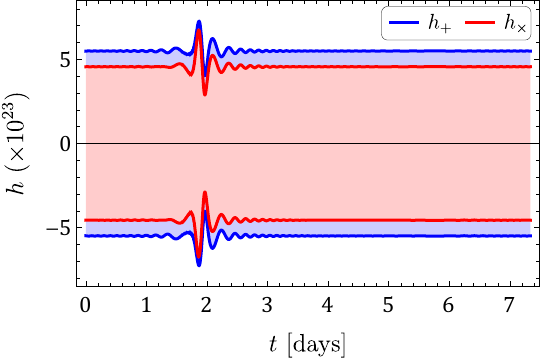}\\
  \includegraphics[width = \columnwidth]{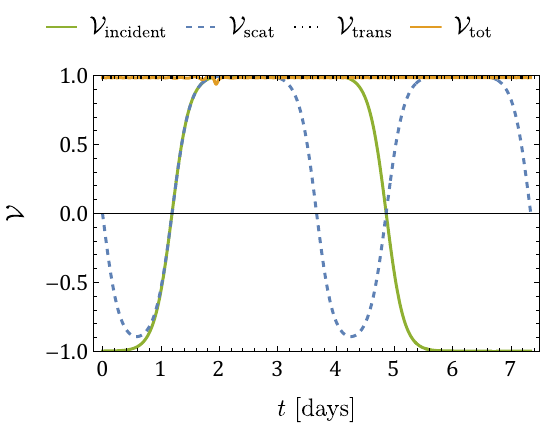}
\caption{Configuration $\iota = 0.12,\, \alpha_1 = 1.40,\, \alpha_2 = 1.00$ \\ \textit{Top }: Envelope of the total wave, for both polarisations, over the course of one orbit in a configuration of high total circular polarisation. Both transmitted $h_{+, \times}$ show a similar amplitude. 
The interference between the scattered and transmitted waves acts as a $\sim 30\%$ to $45\%$ modulation of the wave amplitude, with the modulation being approximately identical to both polarisations in this geometry. Hence, while the interference effect leaves a clear imprint on the polarisations, the total degree of circular polarisation remains approximately constant. 
\textit{Bottom} : Corresponding degree of circular polarisation. Along the 
 orbit $\mathcal{V}_\text{tot} \simeq \mathcal{V}_\text{trans}\simeq 1$, the maximal possible value. }\label{TotalWavesHighV}
\end{figure}

It is shown in Figs.\,\ref{TotalWavesLowV} and \ref{TotalWavesHighV}, for two configurations with respectively  low and  high amount of circular polarisation in the transmitted wave.
In both Figs.\,\ref{TotalWavesLowV} and \ref{TotalWavesHighV}, the total wave signal is dominated by the (constant) amplitude of the  transmitted wave. The interference effect with the scattered wave results in a modulation of the amplitude,  apparent at the point of best alignment.
 The oscillatory pattern arises from interference between the scattered and transmitted waves. Unlike the scattered wave, the transmitted wave's propagation time to the observer varies along the orbit due to changing $d_{\text{LO}}$. 
 Notably, interference oscillations occur when $d_{\text{LO}}$ changes by one wavelength. The peak spacing is larger at quasi-alignment points (extrema of $d_{\text{LO}}(t)$), decreasing as the source moves away from these points. The polarisation modes $h_{+, \times}$ shown in Fig.\,\ref{TotalWavesHighV} have similar amplitude throughout the orbit, for both the transmitted and scattered parts of the waves, leading to an approximately common modulation factor of the amplitudes.

At best alignment, helicity is preserved with $\mathcal{V}_\text{incident} \simeq \mathcal{V}_\text{scat}$, while in anti-alignment (i.e. a configuration lens-source-observer, reached round $t\sim 5$ days),  $\mathcal{V}_\text{incident} \simeq -\mathcal{V}_\text{scat}$, showing that helicity is then reversed by the scattering process. This feature due to the fact that the process is not helicity preserving is general, however the time evolution of $\mathcal{V}_\text{incident}$ and $\mathcal{V}_\text{scat}$ are not universal but configuration dependent. Note that in Figs.\,\ref{TotalWavesLowV} and \ref{TotalWavesHighV}  the degree of polarization of the incident wave $\mathcal{V}_\text{incident}$ and of the scattered one $\mathcal{V}_\text{scat}$ have a similar behaviour. However, the resulting $\mathcal{V}_\text{tot}$ is quite different in the two cases. In the first case indeed, one has $h_{+}>h_{\times}$, hence the relative modulation  due to lensing is larger for the first mode than for the second one. The oscillations in amplitude of the total degree of circular polarisation are due to the oscillations of the single polarisation modes. In the second case, the amplitude of both polarisations is similar, and they are affected in a similar way by lensing. The resulting degree of polarisation is very close to one and shows a small modulation around the point of best alignment. 

\begin{figure}[htb!]
\includegraphics[width = \columnwidth]{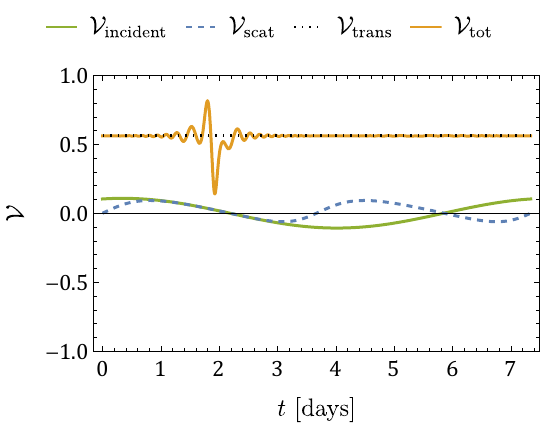}
\caption{Configuration $\iota = 0.15,\, \alpha_1 = 0.16,\, \alpha_2 = 6.12$. Degree of circular polarisation corresponding to the setting of Fig.\,\ref{ScatteredWaves}, with a medium total circularity. Again, we display $\mathcal{V}_\text{incident}$ as the degree of circular polarisation of the incoming wave at the lens, $\mathcal{V}_\text{scat}$ that of the above scattered wave, $\mathcal{V}_\text{trans}$ the (constant) one of the transmitted wave at the observer, and $\mathcal{V}_\text{tot}$ that of the total wave. Both incident and scattered wave have a very low degree of circular polarisation.} 
\label{ScatteredWavesV}
\end{figure}

We stress that $\mathcal{V}_\text{incident}$ and $\mathcal{V}_\text{scat}$ do not need to reach the extremal values of $\pm1$ for a generic orbit, and may actually remain close to 0 at all time. As an example, in Fig.~\ref{ScatteredWavesV} we show a configuration of medium total circularity (it is the same geometry considered in Fig.~\ref{ScatteredWaves} for the scattered wave). 

\section{Discussion and conclusion}\label{sec_discussion}

In this paper we have presented a framework to study lensing of GWs from a black hole, in the deep wave optics regime (i.e. in the case of the wavelength being much larger than the lens Schwarzschild radius),  and keeping full track of the tensorial structure of the wave.  Unlike previous works on GWs lensing in wave optics, we relax the assumption that the wave is a scalar wave.  
We show that the two helicity modes of the wave do not transform in the same way under lensing: the degree of circular polarisation is not preserved by wave optics lensing, unlike what happens in geometric optics and in wave optics when treating the wave as a scalar wave. Our approach shows the limit of the standard treatment of lensing in wave optics where lensing effects are described introducing a common amplification factor for both polarisation modes, and indicates the need to go beyond this approximation to have an accurate description of the lensing signal. 

The formalism used builds up on existing results on phase shifts for spin-2 waves, found in the context of plane wave scattering (see e.g. Ref.~\cite{Dolan:2007ut}). We adapted these results to describe the case of scattering of a wave emitted by a binary system of compact objects, in orbit around a central massive lens, under the assumption that the emitted spherical wave can be considered as a plane wave at the lens scale.    

We presented analytical results for the polarisation modes of the scattered wave, and for the total wave at the observer. We showed that the lensing process is not helicity preserving, unlike what happens in geometric optics. The degree of circular polarisation of a wave is the appropriate basis-independent quantity to highlight this property. 
Although it is not a physical observable in this context, see \cite{Rutherford2022}, we also computed the differential cross section characterising the scattering process and checked that we recover results known in the literature, in some specific limits.

We then turned to the dynamical situation in which the binary system moves around the central lens, on a circular orbit. We illustrated the evolution of the scattered wave with observation time, for an orbital period comparable with the observation time. 

We showed that the total wave received by the observer presents a modulation arising from interference between  the scattered and transmitted components of the wave.\footnote{We illustrated this result for a reference choice of the phase of the transmitted part. As explained in the body of this paper, the computation of the phase distortion of the transmitted component can be done only numerically, and goes beyond the scope of this work. The choice of this phase does not affect quantitatively our findings.}

Hierarchical triple systems of this type are a target for the LISA mission \cite{Toubiana:2020drf, Sberna:2022qbn}. We found that the magnitude of the scattered wave is suppressed with respect to the un-lensed signal as the distance between source and lens increases: sources in disk migration traps around an AGN systems are interesting systems where lensing signals could be sizable. 
Moreover, the relative magnitude of the scattering effect highly depends on the degree of best alignment, given by the inclination parameter $\iota$ of the plane of the outer orbit, with respect to the lens-observer direction. This is summarised by the dimensionless $\rho_\text{max}$ in Eq.~\eqref{relmod2}, which approximates the  maximal value that relative size of the modulation for a given configuration. Notice that this modulation can be large, even in the wave optics regime. We indeed found that the modulation effect can be of the order of $100\%$ for the most favourable geometries and can be detectable by LISA for systems with SNR (in the absence of lensing) sufficiently high, e.g. if triple systems sufficiently close are detected.\footnote{The allowed inclination angle is limited by the assumption of local planarity of the incident wave. When illustrating our results, we have stayed well within the limits of validity of our approximation (meaning that more drastic effects may possibly arise).}

Properly studying lensing effects, accounting for the full tensorial structure of the wave is relevant since matter effects could either be degenerate with tests of general relativity  with low-frequency GW signals, or bias the results of such tests if not properly taken into account. Indeed, some modifications of general relativity also predict low-frequency phase contributions, some already partially constrained by LIGO/Virgo observations~\cite{LIGOScientific:2018dkp,LIGOScientific:2021sio}. LISA has the potential to improve low-frequency constraints by about 8 orders of magnitude (in the flux)~\cite{Barausse:2016eii,Toubiana:2020vtf}, provided that matter effects are absent or accurately modelled.

We note that in our study we have focused on the effect of lensing, assuming that kinematic effects (entering e.g. via the variation of redshift over time see e.g. \cite{Bonvin:2016qxr}) and other line-of-sight-effects are vanishing.  However, for the systems under study, Refs.~\cite{Wong:2019hsq,Randall:2019sab,DOrazio:2019fbq,Yu:2020dlm} found that the Doppler modulation could be measured and used to infer the SMBH mass. If the binary is even closer to the SMBH, its spin can also be measured thanks to the Lens-Thirring effect~\cite{Yu:2020dlm}, and retro-lensing~\cite{Yu:2021dqx} can be significant. Incorporating effects due to the AGN's spin obviously requires going beyond the Schwarzschild spacetime that was considered in this work.
Finally, also note that Refs.~\cite{Hoang:2019kye,Deme:2020ewx,Chandramouli:2021kts} explored other higher order effects (such as the Kozai-Lidov eccentricity oscillations) and found that these could also be detected by LISA in particular configurations.  In a future work, we will review all possible effects affecting the emitted waveform of such a system and analyse the relative importance in various parts of the parameter space. 

\section*{Acknowledgements}
We thank Ruth Durrer and Nicola Tamanini for discussions during different stages of this work. The work of Giulia Cusin, Cyril Pitrou and Jean-Philippe Uzan is supported by CNRS. The work of Giulia Cusin and Martin Pijnenburg is supported by SNSF Ambizione grant \emph{Gravitational wave propagation in the clustered universe}.

\newpage
\appendix

\appendix

\section{Regge-Wheeler variables associated with the incoming wave}\label{matching}

This appendix details the identification of the even and odd Regge-Wheeler potentials associated with the plane wave (\ref{H}) in a system of spherical coordinates centered on the lens. The identification of the even and odd Regge-Wheeler potentials will be asymptotical ($\omega r\gg 1$, which together with $\omega M\ll1$ implies $r\gg M$) to allow the use of the phase shift formalism. To that goal, we decompose the plane wave in spherical components (radial and tangential) in spherical coordinates.  We use the orthonormal $(e_r,e_\theta,e_\phi)$, and the coordinates $(r,\theta,\phi)$ naturally associated with the Cartesian system $(e_x,e_y,e_z)$ attached to the lens and the associated coordinates $(x,y,z)$. We also define the helicity basis
\be
e_\pm = \frac{1}{\sqrt{2}}(e_\theta \mp \ii e_\phi)\,.
\ee
Using the results by Ref.~\cite{Pitrou:2019ifq}, we get the decomposition of the plane wave
in a spherical system of coordinates around the lens, for $p,q = r, \theta, \phi$ 

\be
\label{PitrouPereiraExpansion}
h_{pq} =\sum_{\ell m s} H^{(m)}\,{}_s g^{2m} c_\ell \,{}_s \alpha^{2m}_\ell(kr)\,{}_s Y^{\ell m} n^s_{pq}\cc\,,
\ee
 where the wedge bracket denotes symmetric traceless tensor 
 \be
n^0_{pq} = e^r_{\langle p} e^r_{q \rangle}\qquad n^{\pm 1}_{pq} = e^{\pm}_{\langle p} e^r_{q \rangle}\qquad  n^{\pm 2}_{pq} = e^{\pm}_{\langle p} e^\pm_{q \rangle}\,,
\ee
and  the $c_\ell  = {\rm i}^\ell \sqrt{(4\pi)(2\ell+1)}$ and ${}_s g^{jm}$ are just numerical coefficients that read, from (4.11) of Ref.~\cite{Pitrou:2019ifq},
\be
_0g^{22} = \frac{3}{2}, \quad _{\pm1}g^{22} = \mp \sqrt{3}, \quad _{\pm2}g^{22} = \sqrt{\frac{3}{2}}\,,
\ee
with $_s g^{\ell, -m}=_s g^{\ell m}$.

The radial functions are decomposed into even and odd parity as (here notation with $s \ge 0$)
\be
{}_{\pm s} \alpha^{2m}_\ell = {}_s \epsilon^{2m}_\ell \pm {\rm i} {}_s \beta^{2m}_\ell\,.
\ee
Their explicit expressions are provided in appendix \ref{radial}.  In Eq.~\eqref{PitrouPereiraExpansion}, $_sY^{\ell m}$ are evaluated at $(\theta, \phi)$.

We now want to relate our plane wave~(\ref{PitrouPereiraExpansion}) to the metric decomposition presented in \S.~\ref{Martel}, directly related to Regge-Wheeler variables.  First of all, we recall the relation between even and odd vector harmonics and spin-weighted spherical harmonics which involves the covariant derivative on the unit sphere $D_A$

\begin{align}
Y_A^{\ell m} &= D_A Y^{\ell m} = \frac{1}{r}\sqrt{\frac{\ell(\ell+1)}{2}}\left(-{}_1 Y^{\ell m} e^+_A + {}_{-1} Y^{\ell m} e^{-}_A \right)\,,\\
X_A^{\ell m}& = -\epsilon_A^{\,\,B}D_B Y^{\ell m} \nonumber\\ & =  - \frac{\ii}{r} \sqrt{\frac{\ell(\ell+1)}{2}}\left({}_1 Y^{\ell m} e^+_A + {}_{-1} Y^{\ell m} e^{-}_A\right)\,,
\end{align}
where the $1/r$ rescaling accounts for the fact that we follow \cite{Pitrou:2019ifq} in which $e^\pm_A$ are defined at any $r$, while we introduced $D_A$ on the unit sphere as in \cite{Martel:2005ir}, so that $Y_A^{\ell m}, X_A^{\ell m}$ depend only on $\theta, \phi$.

For even and odd tensor harmonics, we have 
\begin{align}
Y_{AB}^{\ell m} &= D_{\langle A} D_{B \rangle} Y^{\ell m} \nn\\
&=\frac{1}{2r^2}\sqrt{\frac{(\ell+2)!}{(\ell-2)!}}\left({}_2 Y^{\ell m} n^2_{AB} + {}_{-2} Y^{\ell m} n^{-2}_{AB} \right)\,,\\
X_{AB}^{\ell m} &= \frac{\ii}{2r^2}\sqrt{\frac{(\ell+2)!}{(\ell-2)!}}\left({}_2 Y^{\ell m} n^2_{AB} - {}_{-2} Y^{\ell m} n^{-2}_{AB} \right)\,.
\end{align}
Then the radial components of the metric perturbations (\ref{pAB}) are related to the plane wave (\ref{PitrouPereiraExpansion}) as
\be
\label{prrProj}
h_{rr} = \sum_{\ell m} h_{rr}^{\ell m} Y^{\ell m} \,, 
\ee
\be
h_{rA} = h_{Ar} = \sum_{\ell m} h_r^{\ell m}X_A^{\ell m}+j_r^{\ell m}Y_A^{\ell m}\,,
\ee
\be
h_{AB} = \sum_{\ell m} h_2^{\ell m}X_{AB}^{\ell m}+r^2 G^{\ell m}Y_{AB}^{\ell m}+r^2 K^{\ell m}\Omega_{AB}
Y^{\ell m}\,.
\label{pABProj}
\ee
Note that our wave perturbation $h_{ij}$ \eqref{TTwaveMatrix} explicitly has vanishing modes $h_{ta}^{\ell m}, h_{tA}^{\ell m}$.

We verify that under complex conjugation $Y^{\ell m*} = (-1)^m Y^{\ell, -m}$, this relation also holding if one substitutes $Y$ by $Y_{A, B, AB}$ or $X_{A, B, AB}$. As a consequence, one can work with Eq.~\eqref{PitrouPereiraExpansion} first \emph{without} adding the complex conjugate term, thereby extracting the modes \eqref{prrProj}-\eqref{pABProj} of a complexified metric perturbation. The physical (real) metric is then recovered using that, e.g.
\begin{align}
\sum_{\ell m} h_{rr}^{\ell m *} Y^{\ell m*} = &\sum_{\ell m} (-1)^m h_{rr}^{\ell m *} Y^{\ell,- m} \\ = &\sum_{\ell m} (-1)^m h_{rr}^{\ell, -m *} Y^{\ell m}\,, 
\end{align}

so that if $h_{rr}^{\ell m}$ was the expansion coefficient of the complex metric perturbation \eqref{PitrouPereiraExpansion} without c.c., the corresponding coefficient for the real perturbation is $(h_{rr}^{\ell m}+(-1)^m h_{rr}^{\ell, -m*})$. The same holds true for all other coefficients.

We call $(-1)^m h_{rr}^{\ell, -m*}$ the \emph{conjugate mode}, or c.m. for short, which, unlike the complex conjugate, corresponds to the term that has to be added in order to recover the real metric perturbation. For a generic coefficient $a^{\ell,m}$ in \eqref{prrProj}-\eqref{pABProj} :
\be
a^{\ell,m}\cm \ \equiv \ a^{\ell,m} +(-1)^m a^{\ell, -m*}\,.
\ee

Matching Eq.~\eqref{PitrouPereiraExpansion} onto Eqs.~\eqref{prrProj}-\eqref{pABProj} thus gives (recall $m=2$ or $m=-2$ only) 
\begin{align}
\label{hK}
h_{rr}^{\ell m} &= H^{(m)}
\,c_\ell\, {}_0 \epsilon^{2m}_\ell(kr) \cm \,, \\ 
K^{\ell m}&= -\frac{1}{2}h_{rr}^{\ell m} \cm \,,
\end{align}
\begin{align}
\label{jrhr}
j_{r}^{\ell m} &= r\ \sqrt{\frac{3}{2\ell(\ell+1)}} H^{(m)}
 \,c_\ell \,{}_1 \epsilon^{2m}_\ell(kr)  \cm \,, \\ h_{r}^{\ell m} &=r \  \sqrt{\frac{3}{2\ell(\ell+1)}} H^{(m)}
 \,c_\ell \,{}_1 \beta^{2m}_\ell(kr)  \cm \,,
 \end{align}
\begin{align}
r^2 G^{\ell m} &= r^2 \ \sqrt{\frac{6(\ell-2)!}{(\ell+2)!}} H^{(m)}
 \,c_\ell \,{}_2 \epsilon^{2m}_\ell(kr)\,,  \cm \\ \label{h2}h_2^{\ell m} &= r^2 \ \sqrt{\frac{6(\ell-2)!}{(\ell+2)!}} H^{(m)}
 \,c_\ell\, {}_2 \beta^{2m}_\ell(kr) \cm \,,
\end{align}
with all radial functions given in appendix \ref{radial}.

 We can now  explicitly write the quadrupole emission in Regge-Wheeler variables.  We use Eqs.~(\ref{hK})-(\ref{h2}) into (\ref{Psioddfromhr}) and (\ref{Psioddfromhr_even}). Very much like in the scalar partial wave expansion, the gravitational wave partial wave expansion contains spherical Bessel functions and related expressions. It is useful to expand the latter special functions in the asymptotic $kr \gg 1$ limit, to get a result in terms of a superposition of simple spherically in- and outgoing waves. We obtain that the complexified incoming wave (emitted by the binary) around the lens, in the lens frame is described by the following Regge-Wheeler variables:
    \begin{widetext}
\begin{align}
    &\Psi^{\ell m}_\text{even} =\ii\frac{H^{(m)}}{2k}\,\sqrt{6\pi}\sqrt{2\ell+1}\sqrt{\frac{(\ell-2)!}{(\ell+2)!}} \, \bigg((-1)^{\ell+1} e^{- \ii k r}+e^{\ii k r}\bigg) \cm \,, \\
&\Psi^{\ell m}_\text{odd} =- \frac{m}{|m|}\frac{H^{(m)}}{2k}\,\sqrt{6\pi}\sqrt{2\ell+1}\sqrt{\frac{(\ell-2)!}{(\ell+2)!}} \, \bigg((-1)^{\ell+1} e^{- \ii k r}+e^{\ii k r}\bigg) \cm \,.
\end{align}
\end{widetext}

\section{Radial functions}\label{radial}

We list here some useful results, from \cite{Pitrou:2019ifq}, that we used in section \ref{sec_in} and in \ref{matching}. The radial functions ${}_{\pm s} \alpha^{jm}_\ell$ satisfy the following properties 
\be
{}_{-s} \alpha^{jm}_\ell(\nu)={}_{s} \alpha^{j-m}_\ell(\nu)={}_{ s} \alpha^{jm}_\ell(-\nu)\,,
\ee
\be
{}_{s} \alpha^{jm}_\ell={}_{m} \alpha^{js}_\ell\,,
\ee
and
\be
{}_{s} \alpha^{jm}_\ell=(-1)^{\ell-j}{}_{s} \alpha^{\ell m}_j\,.
\ee
We split the function into even and odd part as 
\be
{}_{\pm s} \alpha^{2m}_\ell = {}_s \epsilon^{2m}_\ell \pm {\rm i} {}_s \beta^{2m}_\ell\,,
\ee
where
\be
{}_0 \beta^{2m}_\ell=0\,,
\ee
and
\begin{align}
{}_s \beta^{j, -m}_\ell&=-{}_s \beta^{j, m}_\ell\,,\\
{}_s \epsilon^{j, -m}_\ell&={}_s \epsilon^{j, m}_\ell\,.
\end{align}
In flat space, the radial functions are built as
\be\label{alphaflat}
{}_{s} \alpha^{jm}_\ell(x)=\sum_L {}^{s} C^{m0m}_{\ell L j}j_L(x) i^{L+j-\ell} \sqrt{\frac{4\pi(2L+1)}{(2\ell+1)(2j+1)}}\,,
\ee
in terms of the Gaunt coefficient 
    \begin{align}
{}^{s} C^{m_1 m_2m_3}_{\ell_1\ell_2\ell_3}&\equiv \int d^2\Omega\, \left({}_{s}Y^{\ell_1 m_1}\right)^{*} Y^{\ell_2 m_2}  \left({}_{s}Y^{\ell_3 m_3}\right)\nn\\
&=(-1)^{m_1+s} \sqrt{\frac{(2\ell_1+1)(2\ell_2+1)(2\ell_3+1)}{4\pi}}\nn\\
&\times\left(\begin{array}{ccc}
\ell_1&\ell_2&\ell_3\\
s&0&-s
\end{array}
\right)
\left(\begin{array}{ccc}
\ell_1&\ell_2&\ell_3\\
-m_1&m_2&m_3
\end{array}
\right)\,.
\end{align}
The expressions for the even and odd parity are given in appendix F of \cite{Pitrou:2019ifq}. We list here the functions of interest for us ($x=kr$), and we also mark the scaling for $x \gg 1$
\begin{align}
    {}_0 \epsilon^{2,2}_\ell&=\sqrt{\frac{3(\ell+2)!}{8(\ell-2)!}}\frac{j_{\ell}(x)}{x^2}\propto \frac{1}{x^3}\,,\\
 {}_1 \epsilon^{2,2}_\ell&=\frac{\sqrt{(\ell+2)(\ell-1)}}{2}\frac{1}{x^2}\frac{d}{dx}[xj_{\ell}(x)]\propto \frac{1}{x^2}\,,\\
  {}_1 \beta^{2,2}_\ell&=-\frac{\sqrt{(\ell+2)(\ell-1)}}{2}\frac{j_{\ell}(x)}{x}\propto \frac{1}{x^2}\,,\\
      {}_2 \epsilon^{2,2}_\ell&=\frac{1}{4}\left[j_{\ell}''(x)-j_{\ell}(x)+4\frac{j_{\ell}'(x)}{x}+2\frac{j_{\ell}(x)}{x^2}\right]\propto \frac{1}{x}\,,\\
         {}_2 \beta^{2,2}&=-\frac{1}{2 x^2} \frac{d}{dx}\left[x^2 j_{\ell}(x)\right]\propto \frac{1}{x}\,.
\end{align}

\section{Resummation formulas}\label{Cyril}

We present here resummation formulas that we will use in appendix \ref{app:scattered} to resum the series defining the scattered wave (see section \ref{sec_scatt}). For $|m| = \ell$ and for ${\rm Re}(a) > 0$, we have
\be\label{Magic2}
(1- \cos \theta)^{a-1} {}_s Y^{\ell m}(\theta, \phi) = \sum_{L} {}_s c_L^{\ell m} {}_s Y^{Lm}(\theta,\phi)\,.
\ee
The coefficients are given by (for $m\geq 0$ but possibly $s$ negative) 
\bea\label{cl1}
{}_s c_L^{\ell m} &=&\sqrt{(2\ell+1)(2L +1)} 2^{a-1}\nonumber\\
&&\times\sqrt{\frac{(\ell+m)!(\ell-m)!}{(\ell+s)!(\ell-s)!}\frac{(L+m)!(L-s)!}{(L-m)!(L+s)!}}\nonumber\\
&&\times\frac{\Gamma(m+s+a)\Gamma(L+1-m-a)}{\Gamma(1-a)\Gamma(L+1+m+a)}\,.
\eea
The coefficients when $m<0$ is found from the properties of the spherical harmonics when the $m$ sign is flipped. One finds 
\bea\label{cl2}
{}_s c_L^{\ell m} &=&\sqrt{(2\ell+1)(2L +1)} 2^{a-1}\nonumber\\
&&\times\sqrt{\frac{(\ell+m)!(\ell-m)!}{(\ell+s)!(\ell-s)!}\frac{(L+|m|)!(L+s)!}{(L-|m|)!(L-s)!}}\nonumber\\
&&\times\frac{\Gamma(|m+s|+a)\Gamma(L+1-|m|-a)}{\Gamma(1-a)\Gamma(L+1+|m|+a)}\,.
\eea
The expansion~\eqref{Magic2} is a generalisation of the expansion of $(1- \cos \theta)^{a-1}$ in spherical harmonic, which can be obtained with $s = \ell = m = 0$, and which is obtained from the integral (7.127) of \cite{gradshteyn2007} along with the orthonormality of spherical harmonics. We did not obtain an explicit proof of the expansion~\eqref{Magic2} in the more general case, but we have checked that
\be
{}_s c_L^{\ell m} = \int d^2\Omega\, 
(1- \cos \theta)^{a-1} {}_s Y^{\ell m}(\theta, \phi) \left({}_{s}Y^{Lm}\right)^\star
\ee
on many combinations of $\ell$, $s$, and $L$ with $|m| = \ell$, using the explicit forms of the spin-weighted spherical harmonics.

 \begin{widetext}
\section{Scattered wave: explicit computation}\label{app:scattered}
We want to compute the plus and cross polarisation of the scattered wave of section \ref{sec_scatt}. We need to compute the following objects
\begin{align}
h_+\pm\ii h_{\times}&= \sqrt{6\pi} \ii \frac{e^{\ii kr_*}}{4kr}\, H^{(\pm2)}\sum_\ell \sqrt{2\ell+1} \left(e^{2\ii\delta^{\text{even}}_{\ell}}-e^{2\ii\delta_{\ell}^{\text{odd}}}\right) \,_{\pm2}Y^{\ell \,\pm2}\label{h++ihxSumA}\\
& +\sqrt{6\pi} \ii \frac{e^{\ii kr_*}}{4kr}\, H^{(\mp2)} \sum_\ell \sqrt{2\ell+1}\left(e^{2\ii\delta^{\text{even}}_{\ell}}+e^{2\ii\delta_{\ell}^{\text{odd}}}-2\right) \,_{\pm2}Y^{\ell \,\mp2} \label{h++ihxSumB} \\
&- \sqrt{6\pi} \ii \frac{e^{-\ii kr_*}}{4kr}\, H^{(\mp2)*}\sum_\ell \sqrt{2\ell+1} \left(e^{-2\ii\delta^{\text{even}}_{\ell}}-e^{-2\ii\delta_{\ell}^{\text{odd}}}\right) \,_{\pm2}Y^{\ell \pm2} \\
&- \sqrt{6\pi} \ii \frac{e^{-\ii kr_*}}{4kr}\, H^{(\pm2)*}\sum_\ell \sqrt{2\ell+1} \left(e^{-2\ii\delta^{\text{even}}_{\ell}}+e^{-2\ii\delta_{\ell}^{\text{odd}}}-2\right) \,_{\pm2}Y^{\ell \mp2}\,.\label{h+-ihxSumA}
\end{align}

The summability of the series is discussed in Appendix \ref{sec:DivergingSeries}. 
The sum and difference of phase shifts are given by the following expressions, for  $m=-s$ 
\be\label{deltasoppositem}
\exp\left(2 \ii \delta^{\text{even}}_{L}\right)+\exp\left(2\ii \delta_{L}^{\text{odd}}\right)=2\frac{\Gamma(L-1-2\ii M\omega)}{\Gamma(L+3+2\ii M\omega)}\frac{(L+2)!}{(L-2)!}e^{-\ii \Phi-2\ii M\omega}\,,
\ee
while for $m=s$
\be\label{deltasequalm}
\exp\left(2 \ii \delta^{\text{even}}_{L}\right)-\exp\left(2\ii \delta_{L}^{\text{odd}}\right)=24 \ii M\omega \frac{\Gamma(L-1-2\ii M\omega)}{\Gamma(L+3+2\ii M\omega)}e^{-\ii \Phi-2\ii M\omega}\,.
\ee
We use the resummation formulas in appendix \ref{Cyril} to evaluate the sum over $\ell$  in (\ref{h++ihxSumA}) and (\ref{h++ihxSumB}), i.e. we use 
\be\label{}
(1- \cos \theta)^{a-1} {}_s Y^{\ell m}(\theta, \phi) = \sum_{L} {}_s c_L^{\ell m} {}_s Y^{Lm}(\theta,\phi)\,,
\ee
where the explicit form of the coefficients is given in (\ref{cl1}) and (\ref{cl2}). 
We obtain for $m=-s = \pm 2$
\be\label{csoppositem}
\,_{s}c_L^{\ell m}=\sqrt{(2\ell+1)(2L+1)} 2^{a-1}\frac{(L+2)!}{(L-2)!} \frac{\Gamma(a)}{\Gamma(1-a)}\frac{\Gamma(L-1-a)}{\Gamma(L+3+a)}\,,
\ee
while for $m=s = \pm 2$
\be\label{csequalm}
\,_{s}c_L^{\ell m}=\sqrt{(2\ell+1)(2L+1)} 2^{a-1} \frac{\Gamma(4+a)}{\Gamma(1-a)}\frac{\Gamma(L-1-a)}{\Gamma(L+3+a)}\,,
\ee
where in our context $a=2\ii M\omega$ (we comment on the fact of applying these formulae for that value of $a$ in Sec.~\ref{sec:DivergingSeries}). Using this in our expressions for $h_+ \pm\ii h_\times$ \eqref{h++ihxSumA}-\eqref{h+-ihxSumA}, we obtain
\begin{align}
h_+\pm\ii h_\times = & \sqrt{6\pi}\ii \frac{e^{\ii kr_*}}{4kr} \frac{\Gamma[1-a]}{\sqrt{5}} 2^{2-a} e^{-\ii\Phi-a}\big(1-\cos \theta \big)^{a-1}\bigg\{ H^{(\pm2)} \frac{6a}{\Gamma[4+a]}\, {}_{\pm2}Y^{2\,\pm2}(\theta, \phi) +  H^{(\mp2)} \frac{1}{\Gamma[a]} 
 \,{}_{\pm2}Y^{2\,\mp2}(\theta, \phi)\bigg\}\label{h++ihxMinus2First}
 \\ - & \sqrt{6\pi}\ii \frac{e^{\ii kr_*}}{4kr} 2 H^{(\mp2)} \sum_{L} \sqrt{2L+1}\ {}_{\pm2}Y^{L \, \mp2}\label{h++ihxMinus2}
 \\ -& \sqrt{6\pi}\ii \frac{e^{-\ii kr_*}}{4kr} \frac{\Gamma[1+a]}{\sqrt{5}} 2^{2+a} e^{\ii\Phi+a}\big(1-\cos \theta \big)^{-a-1}\bigg\{ H^{(\mp2)*} \frac{-6a}{\Gamma[4-a]}\, {}_{\pm2}Y^{2\,\pm2}(\theta, \phi) +  H^{(\pm2)*} \frac{1}{\Gamma[-a]}
 \,{}_{\pm2}Y^{2\,\mp2}(\theta, \phi)\bigg\}\label{h+-ihxMinus2First}\\ 
 + & \sqrt{6\pi}\ii \frac{e^{-\ii kr_*}}{4kr} 2 H^{(\pm2)*} \sum_{L} \sqrt{2L+1}\ {}_{\pm2}Y^{L \, \mp2}\,.\label{h+-ihxMinus2}
\end{align}

We note that the sums \eqref{h++ihxMinus2} and \eqref{h+-ihxMinus2} are the leftovers of the $-2$ contributions in respectively \eqref{h++ihxSumB}, \eqref{h+-ihxSumA}, which were there to ensure that the scattered wave polarisations vanish in $a\to0$ limit (i.e. no scattering).  
The limit $a\rightarrow 0$ can be easily taken in (\ref{h++ihxMinus2First}), (\ref{h+-ihxMinus2First}), recalling the $a\to 0$ behaviour of the different $\Gamma[a]$ contributions:
\begin{equation}
    \frac{\Gamma[1-a] 6a}{\Gamma[4+a]} = a +\mathcal{O}(a^2)\,,\qquad
\frac{\Gamma[1-a]}{\Gamma[a]} = a  +\mathcal{O}(a^2)\,.
\label{aExpansionGammas}
\end{equation}

It follows that one ends up with an already vanishing $a\to0$ limit for the resummed terms (if $\theta\neq0$). 
It is therefore expected that the expressions in \eqref{h++ihxMinus2} and \eqref{h+-ihxMinus2} to be also vanishing for any $\theta\neq0$. Indeed, noticing that 
\be
{}_{\pm2}Y^{L,m}(0, 0) = \sqrt{2L+1}\ \frac{\delta_{m, \mp2}}{2\sqrt{\pi}} = {}_{\pm2}Y^{{L,m}*}(0, 0)\,,
\ee
one rewrites the sums in \eqref{h++ihxMinus2},\eqref{h+-ihxMinus2} as 
\begin{align}
& \sum_{L\geq2} \sqrt{2L+1}\ {}_{\pm2}Y^{L \, \mp2}(\theta, \phi) =  \sum_{L\geq2} \sum_{m = -L}^L \sqrt{2L+1}\ \delta_{m, \mp2}\ {}_{\pm2}Y^{L \, m}(\theta, \phi) = \sum_{L\geq2} \sum_{m = -L}^L 2\sqrt{\pi} \ {}_{\pm2}Y^{{L,m}*}(0, 0)\  {}_{\pm2}Y^{L \, m}(\theta, \phi) \nn\\
& = 2\sqrt{\pi} \sum_{L\geq2} \sum_{m = -L}^L {}_{\pm2}Y^{{L,m}*}(0, 0)\  {}_{\pm2}Y^{L \, m}(\theta, \phi) = 2\sqrt{\pi} \ \delta(\theta)\ \delta(\phi)\frac{1}{\sin\theta}\,,
\label{deltaSum}
\end{align}
where we used the completeness relation of spin weighted spherical harmonics to introduce the Dirac deltas \cite{LedesmaSphericalHarmonicTensor}. 

These contributions in the form of a delta function of the forward direction, 
can be traced to the contribution of the incoming plane
wave in the forward direction.

We recall that the expressions \eqref{PhaseShiftMinus} and \eqref{dd1} for the phase shifts are derived in the long wavelength limit ($M\omega\ll1$). We thus expand the $\Gamma$ factors in our summed expressions the same limit, using Eq.~\eqref{aExpansionGammas}. For $\theta\neq0$, one ends up with the approximate expressions
\begin{align}
h_+ \pm \ii h_\times &\simeq  4 \ii\sqrt{\frac{\pi}{5}}\ {}_{\pm2}Y^{2\,\pm2}(\theta, \phi) \bigg\{ \Aa(r) \left(1-\cos\theta\right)^{a-1} H^{(\pm2)} - \Aa^*(r) \left(1-\cos\theta\right)^{-a-1} H^{(\mp2)*} \bigg\} \nonumber\\
&+ 4\ii \sqrt{\frac{\pi}{5}}\ {}_{\pm2}Y^{2\,\mp2}(\theta, \phi) \bigg\{ \Aa(r) \left(1-\cos\theta\right)^{a-1} H^{(\mp2)} - \Aa^*(r) \left(1-\cos\theta\right)^{-a-1} H^{(\pm2)*} \bigg\}   \,,
\end{align}
with the overall $r-$dependent amplitude factor
\be
\Aa(r) = \sqrt{6} \ \frac{e^{\ii kr_*}}{8kr} \ a 2^{1-a} e^{-\ii\Phi-a}\,.
\ee
Recalling that
\begin{equation}
{}_{\pm2}Y^{2\,\pm2}(\theta, \phi) = \frac{1}{2}e^{\pm2\ii\phi}\sqrt{\frac{5}{\pi}}\bigg(\frac{1-\cos\theta}{2}\bigg)^2 \qquad ,\qquad{}_{\pm2}Y^{2\,\mp2}(\theta, \phi) = \frac{1}{2}e^{\mp2\ii\phi}\sqrt{\frac{5}{\pi}}\bigg(\frac{1+\cos\theta}{2}\bigg)^2\,.
\end{equation}

It follows that in $h_+\pm\ii h_\times$, the $\theta\to0$ divergence is only present in the $s=-m$ contributions, but not in the $s=m$ ones.
Extracting the $h_{+, \times}$ polarisations, we obtain the following structure:
\begin{align}
h_+ &\simeq \Aa(r)\ \ii\ (1-\cos\theta)^{a-1} \frac{1+\cos^2\theta}{2} \bigg(H^{(2)} e^{2\ii\phi}+H^{(-2)} e^{-2\ii\phi}\bigg)\cc\,,\\
h_\times &\simeq \Aa(r)\ (1-\cos\theta)^{a-1} \cos\theta \bigg(-H^{(2)} e^{2\ii\phi}+H^{(-2)} e^{-2\ii\phi}\bigg)\cc\,.
\end{align}
Explicitly, this is:
\begin{align}
\label{closedformhplScat_appendix}
h_+ &\simeq \  \frac{2M}{r} \frac{A_\text{in}}{d_\text{SL}}\frac{1}{1-\cos\theta} \frac{1+\cos^2\theta}{2}{}\left(\cos^4\left(\frac{\tilde{\theta}_L}{2}\right)\cos\left(\varphi-2\phi\right)+\sin^4\left(\frac{\tilde{\theta}_L}{2}\right)\cos\left(\varphi+2\phi\right)\right)\,,\\
\label{closedformhxScat_appendix}
h_\times &\simeq \frac{2M}{r} \frac{A_\text{in}}{d_\text{SL}}\frac{1}{1-\cos\theta} \cos\theta \left(\cos^4\left(\frac{\tilde{\theta}_L}{2}\right)\sin\left(\varphi-2\phi\right)-\sin^4\left(\frac{\tilde{\theta}_L}{2}\right)\sin\left(\varphi+2\phi\right)\right)\,,
\end{align}
with the incoming wave amplitude factor $A_\text{in}$ and a phase $\varphi(r, t)$ defined by
\be
A_\text{in}\equiv4 M_c^{5/3} \left(\frac{\omega}{2}\right)^{2/3}, \quad \varphi(r, t) \equiv \omega(t -d_{\text{SL}*}-r_*) -2\tilde{\phi}_L + \Phi - 2M\omega\left(\ln\left(1-\cos\theta\right)-1-\ln2\right)
\label{eq.phaselogA}
\,.
\ee

When the tensorial nature of the GW is restored from these polarisation components, this scattered wave has the structure of the double TT projection
\be\label{Eq.structure}
h_{ij}(t) = \frac{2M}{r(1- \cos \theta)} P_{e_r}\left\{\frac{1}{d_{\rm SL}} P_{\tilde e_r}\left[A_{ij}(t_{\rm ret}) \right]\right\}\,,
\ee
from a quadrupolar source
\be
A_{ij}(t) = A_\text{in}\left[\cos(\omega t)(\tilde e^x_i \tilde e^x_j - \tilde e^y_i \tilde e^y_j) + \sin(\omega t)(\tilde e^x_i \tilde e^y_j + \tilde e^y_i \tilde e^x_j) \right] \,,
\ee
where a TT projector orthogonal to a unit vector is defined by
\be
P_{n}[A_{ij}] = \left(\perp_{i}^k \perp_j^l -  \frac{1}{2} \perp_{ij}  \perp^{kl}\right)A_{kl}\,,\quad \perp_{i}^j  =\delta_i^j - n_i n^j\,.
\ee
The first TT projection in \eqref{Eq.structure} with $P_{\tilde e_r}$ is the projection of the quadrupolar source onto the plane orthogonal to the radiated direction $\tilde e_r$, with a spherical wave decay $\propto 1/d_{\rm SL}$, which is evaluated at the lens position. The second TT projection with $P_{e_r}$ due to scattering, is in the plane orthogonal to the scattered direction $e_r$, and it is weighted by the $2M/(1- \cos \theta)$ factor with a spherical wave decay $\propto 1/r$. As seen in Eq.~\eqref{eq.phaselogA}, the source is evaluated at a retarded time $t_{\rm ret}$ which is different from the naive $t - d_\text{SL}-r$, not only because tortoise coordinates must be used, hence $d_\text{SL*}$ and $r_*$ replace $d_\text{SL}$ and $r$, but also due to a further offset as in \eqref{eq.phaselogA}.
 \end{widetext}

\section{Summability of the multipole series}
\label{sec:DivergingSeries}
We now discuss the convergence and the summation of the multipole series (\ref{hpmReconstructionSpherical}) defining plus and cross polarisations after the scattering 
\be
h_+\pm \ii h_{\times}=\sum_{\ell m} \ _{\pm2}Y^{\ell m} h^{(\pm 2)}_{\ell m}\,.
\ee
with multipoles given in (\ref{hlm}). 

We start by considering the high-$\ell$ behaviour of $_{\pm2}Y^{\ell m}$. For this we exploit the asymptotic behaviour of the Wigner $d^\ell _{m s}$ function, as well as the relation between $d^\ell _{m s}$ and $_{s}Y^{\ell m}$. The latter is expressed as follows:
\be
_sY^{\ell m}(\theta, \phi) = \sqrt{\frac{2\ell+1}{4\pi}}\, e^{\ii m\phi} \,(-1)^s\, d^\ell _{m,\, -s}(\theta)\,.
\label{YdRelation}
\ee
The asymptotic behaviour of $d^\ell _{m,\, -s}$ is given e.g. in \cite{Rowe}.
In the range $\theta\in[0, \pi]$, we have that for $\ell\to\infty$, and $m\approx -s\approx \frac{1}{2}(m-s)\ll\ell$
\begin{widetext}
\be
d^\ell _{m,\, -s}(\theta)\to
    \begin{cases}
        J_{-s-m}\bigg(\theta \sqrt{\ell^2-\frac{1}{4}(m-s)^2}\bigg) & \text{if } \theta \leq \frac{\pi}{2}\\
        (-1)^{\ell-m}\ J_{m-s}\bigg((\theta-\pi) \sqrt{\ell^2-\frac{1}{4}(m+s)^2}\bigg)  & \text{if } \theta > \frac{\pi}{2}
    \end{cases}\quad,
\ee
\end{widetext}
with $J_n$ being the Bessel function of order $n$. 
The  Bessel functions $J_n(\ell)$ at large $\ell$ scales as $\ell^{-1/2}\cos(\ell)$ (see e.g.\,,\cite{Sekeljic}). Therefore, the asymptotic scaling in $\ell$ of $_{s}Y^{\ell m}$ has dominant term $\propto \ell^{1/2}\ell^{-1/2}\cos(\ell) = \cos(\ell)$, i.e. is an oscillating function with constant amplitude.

We need now to consider the high-$\ell$ behaviour of the coefficients in Eq.~\eqref{hlm}. 

In this context, we note that even and odd phase shifts tend to become more and more identical at high $\ell$, and that their respective contribution is either summed or subtracted in the $m$ partial sum of the full $m, \ell$ multipole sum, depending if the considered partial sum has $m = s$, or $m= -s$, for fixed $s=\pm2$ mode. 

Considering only the lower orders in $M\omega$ for the phase shifts, it appears that the partial sum with $m=s$ sees the even and odd contributions to cancel each better and better with increasing $\ell$ (i.e. these are the terms with $\pm \frac{m}{|m|} = -1$ in Eq.~\eqref{hlm}). Indeed, we have
\begin{widetext}
\be
(e^{2 \ii \delta_{\ell }^{\text{even}}}-1)-(e^{2 \ii \delta_{\ell }^{\text{odd}}}-1) = e^{2 \ii \delta_{\ell }^{\text{even}}}-e^{2 \ii \delta_{\ell }^{\text{odd}}} = 24 \ii M\omega \frac{(\ell-2)!}{(\ell+2)!} +\mathcal{O}\big((M\omega)^2\big)
\ee
\end{widetext}
The $\ell^{-4}$ scaling of this term apparently removes the divergence in these particular partial sums.
On the other hand, in the partial sums with $m= -s$, the divergence of the even and odd subcomponents adds, and the resulting partial sums remain divergent.
The full multipole sum, being the sum of the two partial sums with $m = \pm s$, remains therefore divergent.

It has been illustrated in Ref.~\cite{Rutherford2022} that the simpler problem of Rutherford scattering and of scattering of a scalar wave off a Schwarzschild black hole also typically gives rise to a divergent multipole series for the scattering amplitude. This is a mathematical artifact, due to the improper taking of asymptotic limits within individuals multipoles before summing them. For the case of Rutherford, one can indeed compute the exact solution in real space, and get a closed form for the scattering amplitude.  When working in multipole space, one can obtain a closed form for the series by using summation formulas extended outside their range of validity, just as has been done here using $a = 2\ii M\omega$ in (\ref{cl1}) and (\ref{cl2}).  For Rutherford and for the scalar wave case, this closed form has remarkably been shown to be unique for any converging non standard summation that fulfils criteria of regularity, linearity and stability, such as Cesàro summation. 

Cesàro summation is a (regular, linear and stable) summation technique that can average out the oscillations that are preventing some series from converging.
We briefly introduce it here. Further details can be found in the classic reference \cite{Hardy}. 
For any series with terms $a_j$, we introduce the family of $(C, \alpha)$ Cesàro summations of the series, for $\alpha \in \mathbb{N}$, as  
\be
\lim_{n\to\infty}\sum_{j=0}^n \frac{\binom{n}{j}}{\binom{n+\alpha}{j}}a_j\,.
\ee
The case $\alpha = 0$ corresponds to ordinary summation, while the case $\alpha =1$ is often simply called Cesàro summation. By construction, Cesàro summation is an average of the ordinary partial sums. 
When solving the problem numerically, one has to truncate these sums at some maximal $n$ value. Numerically, partial sums of Cesàro summation of multipoles series was shown to provide results in agreement with the expected closed form apart from the $\theta=0$ direction, in the context of Rutherford scattering see \cite{Rutherford2022}. While, unlike the case of Rutherford, we do not show a unicity result of Eqs.~\eqref{closedformhplScat_appendix}-\eqref{closedformhxScat_appendix} being the closed forms of any converging non standard summation that fulfils criteria of regularity, linearity and stability, it is natural to conjecture this property to hold.

\begin{figure*}[!htb]
\centering
   \includegraphics[width=\columnwidth]{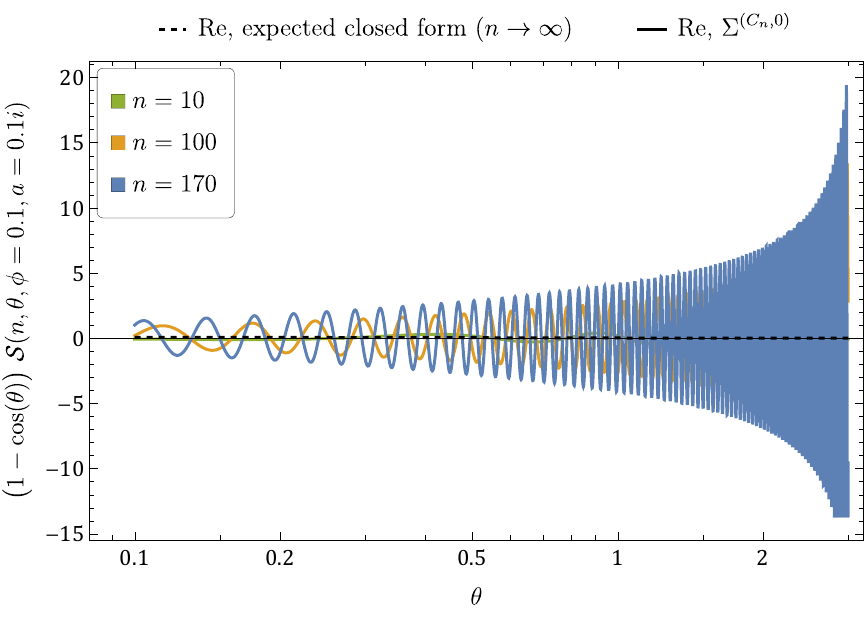}
   \label{RePlotC0}
   \includegraphics[width=\columnwidth]{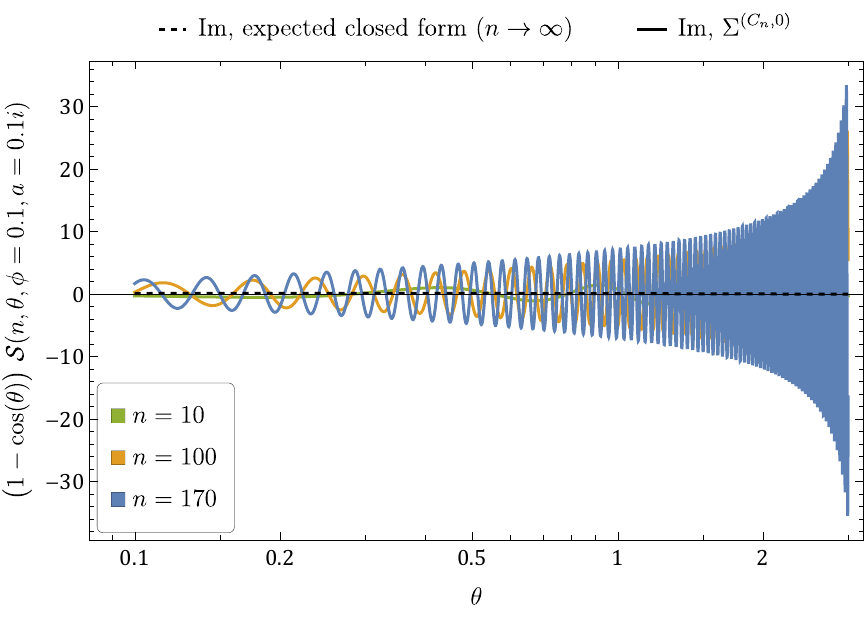}
   \label{ImPlotC0}\\
\includegraphics[width=\columnwidth]{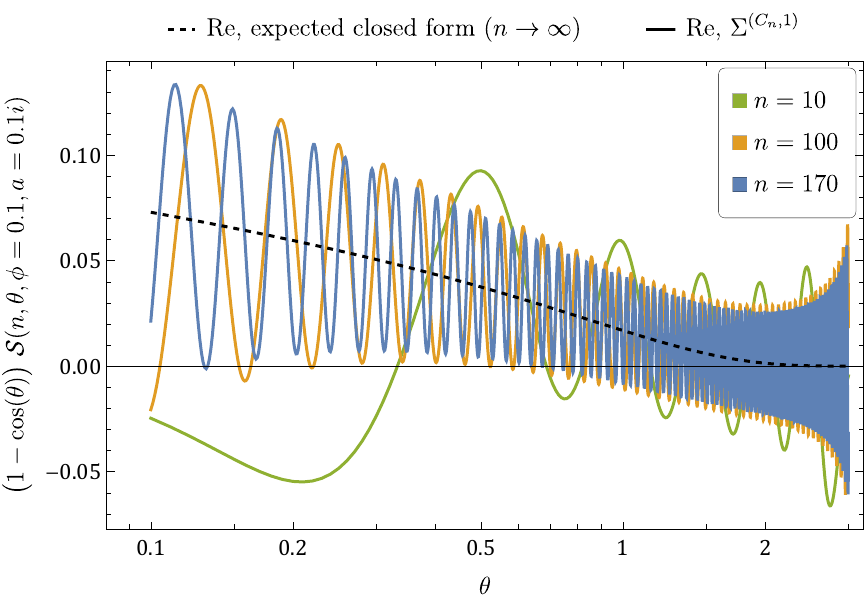}
   \label{RePlotC1}
   \includegraphics[width=\columnwidth]{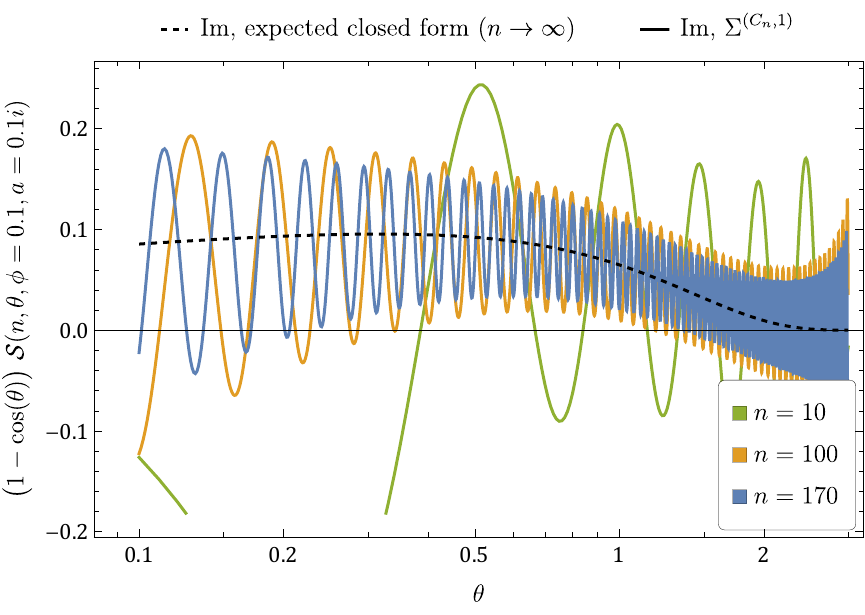}
   \label{ImPlotC1}
\includegraphics[width=\columnwidth]{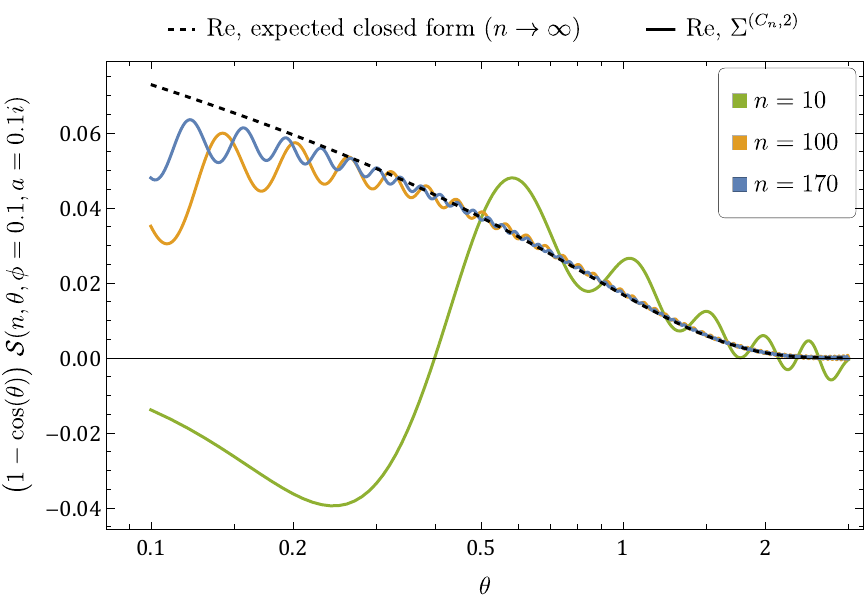}
\label{RePlotC2}
   \includegraphics[width=\columnwidth]{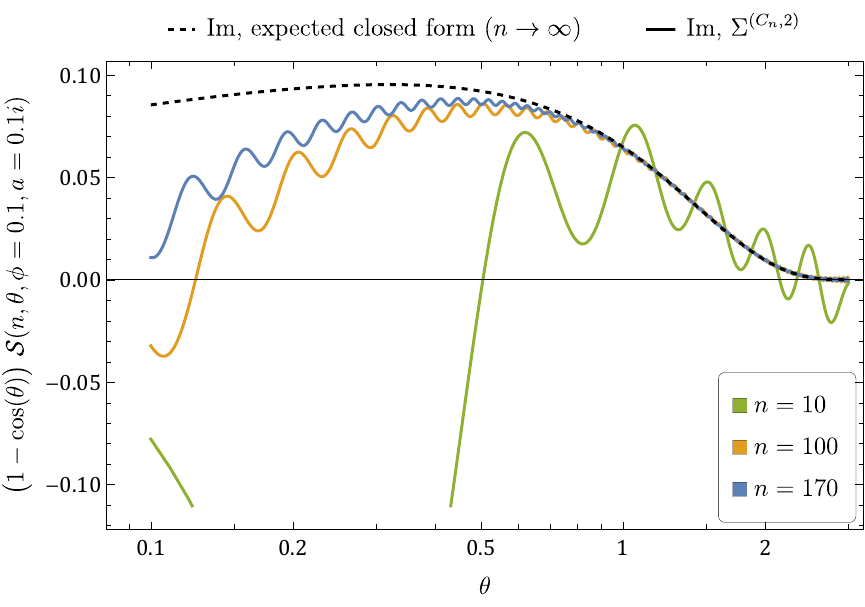}
   \label{ImPlotC2}       
\caption{Real (\textit{left column}) and imaginary (\textit{right column}) parts of the numerical truncated Cesàro summation of series \eqref{sigma}, each time for various maximal values of the summation order $n$. On the first row is the $\alpha = 0$ case of Cesàro summation, i.e. ordinary partial sums, while on the second and third rows are the $\alpha =1,2$ truncated Cesàro summations.}
\label{PlotsSums}
\end{figure*}

In view of investigating this numerically with Cesàro summation in mind, we turn to our $m=-s$ summations of interest, e.g. \eqref{h++ihxSumB} :

\begin{widetext}
\begin{equation}
\mathcal{S}(n\to\infty, \theta, \phi , a)\equiv\sum_{\ell = 2}^{n\to\infty}\left(-2+2e^{-a-\ii\Phi(a)}\frac{\Gamma[\ell-1-a]}{\Gamma[\ell+3+a]} \frac{(\ell+2)!}{(\ell-2)!}\right) \sqrt{2\ell+1}{}_{-2}Y_{\ell,\ 2}(\theta, \phi)\,,
\label{DivergingSeries}
\end{equation}
\end{widetext}
where we used Eq.~ \eqref{deltasoppositem} to sum phase shift exponentials. We shift the summation index by two to rewrite the partial sum as a sum starting at $\ell =0$ (stability property), defining 
\begin{widetext}
\be
\mathcal{S}(n, \theta, \phi , a) = \sum_{\ell = 0}^n \left(-2+2e^{-a-\ii\Phi(a)}\frac{\Gamma[\ell+1-a]}{\Gamma[\ell+5+a]} \frac{(\ell+4)!}{\ell!}\right) \sqrt{2\ell+5} {}_{-2}Y_{\ell+2,\ 2}(\theta, \phi).
\label{sigma}
\ee
\end{widetext}
As $n\to\infty$, we remind that the expected closed form of this series is (see appendix \ref{app:scattered})
\be\label{closed}
\frac{\Gamma[1 - a]}{\sqrt{5} \Gamma[a]} 2^{2 -a} e^{-a - \ii\Phi(a)}\big(1 - \cos \theta\big)^{-1+a}  {}_{-2}Y_{2,\ 2}(\theta, \phi)\,,
\ee
with a $\big(1 - \cos(\theta)\big)^{-1}$ divergence as $\theta\to0$ ($a$ being purely imaginary). We now replace the ordinary partial sums in (\ref{sigma}) with Cesàro partial sums: the Cesàro averaging process averages out the ever growing oscillations of the partial sums.  We illustrate numerically in Fig.~\ref{PlotsSums} the attempt of summing $\mathcal{S}(n, \theta, \phi , a)$ until various finite maximal $n$ values, and replacing the ordinary partial sum with $(C, \alpha = 0, 1, 2)$ partial sums, which we denote by $\sum^{(C_n, \alpha)}$. In each case we rescale the summed result by a $\big(1 - \cos(\theta)\big)$ factor for visual convenience, and we plot the result against the closed form (\ref{closed}). While the sum never converges in the sense of ordinary summation, it numerically converges in the sense of Cesàro. The same holds for the other $s = -m$ sum \eqref{h+-ihxSumA}.

\section{Cross section: comparison with literature }\label{app_cross}

The cross section for scattering of gravitational waves off Schwarzschild black holes has been computed already in in 70s, see e.g. Ref.~\cite{Peters1976}, where both the cross section for unpolarised and purely $+$ or $\times$ polarised incoming waves are given.

For a circularly polarised incoming wave, which in our case reduces to $\tilde{\theta}_L = 0$, our cross section (\ref{FulldsigmadOmega})  reduces to 
\be
\frac{\dd\sigma}{\dd\Omega} = M^2 \frac{\cos^8(\theta/2)+\sin^8(\theta/2)}{\sin^4(\theta/ 2)}\,.
\label{unpolCS}
\ee
This can equivalently be written as 
\be
\frac{\dd\sigma}{\dd\Omega} = M^2 \frac{1}{\sin^4(\theta/2)}\left(\cos^2\theta+\frac{1}{8}\sin^4\theta\right)\,,
\ee
and it agrees with the result of  \cite{Peters1976} for an \emph{unpolarised} wave.  Our cross section in the limit $\tilde{\theta}_L\to\pi/2$ is that of a purely $h_+$ polarised incoming wave. In this limit, after a few simplification we can check that it agrees  with the result of Peters for purely $+$ incoming wave, i.e.
\be
\frac{1}{\sin^4(\theta/2)}\Big(\cos^2\theta+\frac{1}{4}\sin^4\theta\cos^2(2\phi)\Big)\,.
\label{Petersh+only}
\ee
Peters also gives a formula for a purely $\times$ incoming plane wave, and finds
\be
\frac{1}{\sin^4(\theta/2)}\Big(\cos^2\theta+\frac{1}{4}\sin^4 \theta \sin^2(2\phi)\Big)\,.
\label{Petershxonly}
\ee
Our choice of basis in the polarisation plane does not allow to have a purely $\times$ incoming wave. Indeed the only incoming wave with zero circular polarisation that we have is that of a binary seen edge on, with a purely $+$ emission. To obtain a purely $\times$ incoming wave, one should take the configuration of an $h_\times$ only incoming wave (i.e. $\tilde{\theta}_L\to\pi/2$) and rotate the polarisation plane basis (on the lens) by $-\pi/4$. The rotation axis (source-lens axis) is left unchanged by this rotation, so the $\theta$ angle remains untouched. The new angle $\phi$ is however rotated by $\pi/4$ with respect to the old one. Thus, re-expressing the cross section in this new polarisation basis, one has to substitute $\phi \to \phi -\pi/4 $ in \eqref{Petersh+only}. This self-consistently gives \eqref{Petershxonly}, as expected.
Averaging the result for the pure $h_+$ and $h_\times$ incoming wave as \cite{Peters1976}, we indeed obtain their claimed result for an unpolarised incoming wave, i.e. \eqref{unpolCS}.

\FloatBarrier
\section{Triple system: parametrising the time-dependence}\label{triple}

We now want to look at emission of a wave from source orbiting around a fixed lens (AGN), with a hypothetical fixed observer. The motion of the source is on 3-dimensional orbit. Fig.\,\ref{MovingScheme}, represents a snapshot at a fixed time. 

We now wish to parametrise the source's circular motion around the lens. 
For that, we introduce a further set of Cartesian orthonormal basis vectors $(\bar{e}_x, \bar{e}_y, \bar{e}_z)$ that are attached to the observer-lens rest frame, centered on the lens. For simplicity, we fix $\bar{e}_x$ to be in the observer direction, and fix the time origin of the circular motion such that the source is initially along $\bar{e}_y$. With that, the orbital trajectory of the source is expressed as

\be
\bar{\br}_S(t) = d_\text{SL} \begin{pmatrix} - \cos\iota \sin(\Omega t)\\  \cos(\Omega t) \\ \sin\iota \sin(\Omega t)\end{pmatrix}\quad (\text{w.r.t.}\ \bar{e}_{x, y, z})\,,
\label{SourceTrajectory}
\ee
which is identical to the parametrisation chosen in \cite{DOrazio:2017ssb, Sberna:2022qbn} (for $\phi_\bullet\to0$ in the former reference).
The angle $\iota$ corresponds to the inclination of the orbital plane with respect to the observer-lens axis and is a free parameter.

Recall that in the earlier static setting, the set of basis vectors $(e_x, e_y, e_z)$ were constructed from the source-attached basis vectors $(\Tilde{e}_x, \Tilde{e}_y, \Tilde{e}_z)$, the former being the set of spherical coordinates unit vectors in the Cartesian basis given by the latter. We keep the same construction here, with the extra hypothesis of axial parallelism: the triad $(\Tilde{e}_x, \Tilde{e}_y, \Tilde{e}_z)$ is simply translated along the source motion, but sees its orientation preserved, i.e. scalar products between $\Tilde{e}_i$ and $\bar{e}_i$ basis vectors (fixed) are constants over time. This neglects any precession of the source's internal angular momentum. Vectors $(\Tilde{e}_x, \Tilde{e}_y, \Tilde{e}_z)$ are therefore uniquely defined by three Euler angles $\alpha_{1, 2, 3}$, which rotate $(\bar{e}_x, \bar{e}_y, \bar{e}_z)$ into $(\Tilde{e}_x, \Tilde{e}_y, \Tilde{e}_z)$. So, in terms of $\bar{e}_{x, y, z}$ 
\begin{widetext}
\be
\Tilde{e}_z \equiv \begin{pmatrix} \sin\alpha_1\cos\alpha_2 \\ \sin\alpha_1\sin\alpha_2  \\ \cos\alpha_1 \end{pmatrix},\,\quad \Tilde{e}_x \equiv \begin{pmatrix} \cos\alpha_1\cos\alpha_2\cos\alpha_3-\sin\alpha_2\sin\alpha_3 \\ \cos\alpha_1\sin\alpha_2\cos\alpha_3 +\cos\alpha_2\sin\alpha_1\\ -\sin\alpha_1\cos\alpha_3 \end{pmatrix} ,\,\quad \Tilde{e}_y \equiv \Tilde{e}_z\times \Tilde{e}_x\,,
\label{SourceBasis}
\ee
\end{widetext}
where $\alpha_{1, 2, 3}$ are free parameters. While $\alpha_{1,2}$ have a physical meaning associated with the source's internal angular momentum (parallel to $ \tilde{e}_z$), $\alpha_3$ merely fixes the orientation of $\tilde{e}_{x, y}$.

With the source orbital motion and axial parallel translation of $(\Tilde{e}_x, \Tilde{e}_y, \Tilde{e}_z)$, the $(\tilde{\theta}_{L}, \tilde{\phi}_{L}, \tilde{\theta}_{\text{O}}, \tilde{\phi}_{ \text{O}})$ angular positions of lens and observer become functions of time.
Similarly, $(e_x, e_y, e_z)$ rotates as a function of time, and so $(\theta,\phi)$ become functions of time.

From the simple orbital motion \eqref{SourceTrajectory}, the angle between the source-lens axis and the lens-observer axis is easily computed to be
\be
\theta(t) = \arccos(\cos\iota \sin(\Omega t))\,.
\ee
The observer azimuthal angle with respect to $e_{x, y, z}$ is more involved. We find 
\begin{widetext}
			\begin{align}
				&\tan\big(\phi(t)\big) = \\
				& \frac{\sin \alpha_1 \sin \iota \sin (\Omega t) \sin \alpha_2-\cos \alpha_1 \cos (\Omega t)}{\sin \alpha_1 \cos \alpha_2 \big[\frac{1}{2} \cos ^2\iota  \cos (2
					\Omega t)-\frac{1}{4} \cos (2 \iota )+\frac{3}{4}\big]+\cos \iota \sin(\Omega t) \big[ \cos \alpha_1 \sin \iota \sin (\Omega t)+ \sin \alpha_1 \cos (\Omega t)\sin
					\alpha_2\big]}\nn,
			\end{align}
				while $	\tilde{\theta}_L(t)$ is computed from $\tilde{e}_z\cdot e_z$, giving
			\be
			\tilde{\theta}_L(t) = \arccos\Big(\sin (\Omega t) \Big[\sin \alpha_1 \cos \iota \cos \alpha_2-\cos \alpha_1 \sin \iota\Big]-\sin \alpha_1 \cos (\Omega t)\sin \alpha_2\Big) \,.
			\ee
\end{widetext}
Further
\be
\tan\left(\tilde{\phi}_L(t)\right) = \frac{\chi_1\cos{\alpha_3}+ \chi_2\sin{\alpha_3}}{\chi_1\sin{\alpha_3}-\chi_2\cos{\alpha_3}}\,,
\ee
for 
\be
\chi_1\equiv \cos(\Omega t)\cos{\alpha_2}+\sin(\Omega t)\cos{\iota}\sin{\alpha_2}\,,
\ee
and
\begin{align}
    \chi_2 \equiv &- \cos(\Omega t)\cos{\alpha_1}\sin{\alpha_2} \nn\\
    &+\sin(\Omega t)\left(\sin{\iota}\sin{\alpha_1}+\cos{\iota}\cos{\alpha_1}\cos{\alpha_2}\right)\,.
\end{align}
As for the transmitted wave, we consider the incident wave emitted by the source, evaluated at observer position.
From \eqref{SourceBasis}, the angular position of the observer as seen from the source's Cartesian $\tilde{e}_{x, y, z}$ is given by 
\be
\cos(\tilde{\theta}_\text{O}) =\sin\alpha_1\cos\alpha_2\,.
\label{thetaObsFar}
\ee
\be
\tan(\tilde{\phi}_\text{O, far})= \frac{-\cos\alpha_1 \cos\alpha_2 \sin\alpha_3 - \sin\alpha_2\cos\alpha_3}{\cos\alpha_1 \cos\alpha_2\cos\alpha_3  - \sin\alpha_2\sin\alpha_3 }
\label{phiObsFar}
\ee
This position is independent of time in the far observer
limit.
Finally, from the orbital motion parametrised by \eqref{SourceTrajectory}, we find from $(\tilde{e}_{\tilde{\theta}}\cdot e_\theta)|_\text{O}  = (\tilde{e}_{\tilde{\phi}}\cdot e_\phi)|_\text{O}$ 
\begin{widetext}
    \be \cos(\eta) = 
-\frac{2 (\cos(\Omega t) \sin\alpha_1 \sin\alpha_2 + \cos\alpha_1 \sin\iota \sin(\Omega t))}{\sqrt{\cos^2\alpha_1 \cos^2\alpha_2 + \sin^2\alpha_2} \sqrt{3 + \cos(2 \Omega t) - 2 \cos(2 \iota) \sin^2(\Omega t)}}
\ee
 \end{widetext}
The notation used for all geometrical concepts is summed up in Table \ref{TableBig}.
Due to finite travel time of the GW signal, the scattering geometry should be evaluated at retarded time.

\FloatBarrier

\bibliographystyle{bib-style}
\bibliography{myrefs}
\end{document}